\documentclass[twocolumn,trackchanges]{aastex7}

\usepackage{CJK}
\newcommand{\corr}[1]{{\color{black}#1}}

\usepackage{xcolor}

\begin{document}

\title{Spectral Diversity in Type Ibn Supernovae and the Large Host Offset of SN~2024acyl}



\newcommand{\LCO}{\affiliation{Las Cumbres Observatory, 6740 Cortona Drive, Suite 102, Goleta, CA 93117-5575, USA}}
\newcommand{\UCSB}{\affiliation{Department of Physics, University of California, Santa Barbara, CA 93106-9530, USA}}
\newcommand{\KITP}{\affiliation{Kavli Institute for Theoretical Physics, University of California, Santa Barbara, CA 93106-4030, USA}}
\newcommand{\UCD}{\affiliation{Department of Physics and Astronomy, University of California, Davis, 1 Shields Avenue, Davis, CA 95616-5270, USA}}
\newcommand{\WIS}{\affiliation{Department of Particle Physics and Astrophysics, Weizmann Institute of Science, 76100 Rehovot, Israel}}

\newcommand{\OKC}{\affiliation{Oskar Klein Centre, Department of Astronomy, Stockholm University, Albanova University Centre, SE-106 91 Stockholm, Sweden}}
\newcommand{\OAPD}{\affiliation{INAF-Osservatorio Astronomico di Padova, Vicolo dell'Osservatorio 5, I-35122 Padova, Italy}}
\newcommand{\Caltech}{\affiliation{Cahill Center for Astronomy and Astrophysics, California Institute of Technology, Mail Code 249-17, Pasadena, CA 91125, USA}}
\newcommand{\GSFC}{\affiliation{Astrophysics Science Division, NASA Goddard Space Flight Center, Mail Code 661, Greenbelt, MD 20771, USA}}
\newcommand{\UMD}{\affiliation{Joint Space-Science Institute, University of Maryland, College Park, MD 20742, USA}}
\newcommand{\UCB}{\affiliation{Department of Astronomy, University of California, Berkeley, CA 94720-3411, USA}}
\newcommand{\TTU}{\affiliation{Department of Physics, Texas Tech University, Box 41051, Lubbock, TX 79409-1051, USA}}
\newcommand{\STScI}{\affiliation{Space Telescope Science Institute, 3700 San Martin Drive, Baltimore, MD 21218-2410, USA}}
\newcommand{\UT}{\affiliation{University of Texas at Austin, 1 University Station C1400, Austin, TX 78712-0259, USA}}
\newcommand{\IoA}{\affiliation{Institute of Astronomy, University of Cambridge, Madingley Road, Cambridge CB3 0HA, UK}}
\newcommand{\QUB}{\affiliation{Astrophysics Research Centre, School of Mathematics and Physics, Queen's University Belfast, Belfast BT7 1NN, UK}}
\newcommand{\IPACSSC}{\affiliation{Spitzer Science Center, California Institute of Technology, Pasadena, CA 91125, USA}}
\newcommand{\IPAC}{\affiliation{IPAC, California Institute of Technology, 1200 East California Boulevard, Pasadena, CA 91125, USA}}
\newcommand{\JPL}{\affiliation{Jet Propulsion Laboratory, California Institute of Technology, 4800 Oak Grove Dr, Pasadena, CA 91109, USA}}
\newcommand{\Southampton}{\affiliation{Department of Physics and Astronomy, University of Southampton, Southampton SO17 1BJ, UK}}
\newcommand{\LANL}{\affiliation{Space and Remote Sensing, MS B244, Los Alamos National Laboratory, Los Alamos, NM 87545, USA}}
\newcommand{\Tsinghua}{\affiliation{Physics Department and Tsinghua Center for Astrophysics, Tsinghua University, Beijing, 100084, People's Republic of China}}
\newcommand{\NAOC}{\affiliation{National Astronomical Observatory of China, Chinese Academy of Sciences, Beijing, 100101, People's Republic of China}}
\newcommand{\YNAO}{\affiliation{Yunnan Observatories (YNAO), Chinese Academy of Sciences (CAS), Kunming, 650216, People's Republic of China}}
\newcommand{\ICEY}{\affiliation{International Centre of Supernovae, Yunnan Key Laboratory, Kunming 650216, People's Republic of China}}
\newcommand{\UAS}{\affiliation{School of Astronomy and Space Science, University of Chinese Academy of Sciences, Beijing 100049, People's Republic of China}}
\newcommand{\Itagaki}{\affiliation{Itagaki Astronomical Observatory, Yamagata 990-2492, Japan}}
\newcommand{\Einstein}{\altaffiliation{Einstein Fellow}}
\newcommand{\Hubble}{\altaffiliation{Hubble Fellow}}
\newcommand{\CfA}{\affiliation{Center for Astrophysics \textbar{} Harvard \& Smithsonian, 60 Garden Street, Cambridge, MA 02138-1516, USA}}
\newcommand{\UA}{\affiliation{Steward Observatory, University of Arizona, 933 North Cherry Avenue, Tucson, AZ 85721-0065, USA}}
\newcommand{\MPIA}{\affiliation{Max-Planck-Institut f\"ur Astrophysik, Karl-Schwarzschild-Stra\ss{}e 1, D-85748 Garching, Germany}}
\newcommand{\DSFP}{\altaffiliation{LSSTC Data Science Fellow}}
\newcommand{\HCO}{\affiliation{Harvard College Observatory, 60 Garden Street, Cambridge, MA 02138-1516, USA}}
\newcommand{\Carnegie}{\affiliation{Observatories of the Carnegie Institute for Science, 813 Santa Barbara Street, Pasadena, CA 91101-1232, USA}}
\newcommand{\TAU}{\affiliation{School of Physics and Astronomy, Tel Aviv University, Tel Aviv 69978, Israel}}
\newcommand{\Edinburgh}{\affiliation{Institute for Astronomy, University of Edinburgh, Royal Observatory, Blackford Hill EH9 3HJ, UK}}
\newcommand{\Birmingham}{\affiliation{Birmingham Institute for Gravitational Wave Astronomy and School of Physics and Astronomy, University of Birmingham, Birmingham B15 2TT, UK}}
\newcommand{\Bath}{\affiliation{Department of Physics, University of Bath, Claverton Down, Bath BA2 7AY, UK}}
\newcommand{\CTIO}{\affiliation{Cerro Tololo Inter-American Observatory, National Optical Astronomy Observatory, Casilla 603, La Serena, Chile}}
\newcommand{\Potsdam}{\affiliation{Institut f\"ur Physik und Astronomie, Universit\"at Potsdam, Haus 28, Karl-Liebknecht-Str. 24/25, D-14476 Potsdam-Golm, Germany}}
\newcommand{\INPE}{\affiliation{Instituto Nacional de Pesquisas Espaciais, Avenida dos Astronautas 1758, 12227-010, S\~ao Jos\'e dos Campos -- SP, Brazil}}
\newcommand{\UNC}{\affiliation{Department of Physics and Astronomy, University of North Carolina, 120 East Cameron Avenue, Chapel Hill, NC 27599, USA}}
\newcommand{\Ohio}{\affiliation{Astrophysical Institute, Department of Physics and Astronomy, 251B Clippinger Lab, Ohio University, Athens, OH 45701-2942, USA}}
\newcommand{\AAS}{\affiliation{American Astronomical Society, 1667 K~Street NW, Suite 800, Washington, DC 20006-1681, USA}}
\newcommand{\MMT}{\affiliation{MMT and Steward Observatories, University of Arizona, 933 North Cherry Avenue, Tucson, AZ 85721-0065, USA}}
\newcommand{\Geneva}{\affiliation{ISDC, Department of Astronomy, University of Geneva, Chemin d'\'Ecogia, 16 CH-1290 Versoix, Switzerland}}
\newcommand{\IUCAA}{\affiliation{Inter-University Center for Astronomy and Astrophysics, Post Bag 4, Ganeshkhind, Pune, Maharashtra 411007, India}}
\newcommand{\CMU}{\affiliation{Department of Physics, Carnegie Mellon University, 5000 Forbes Avenue, Pittsburgh, PA 15213-3815, USA}}
\newcommand{\NAOJ}{\affiliation{Division of Science, National Astronomical Observatory of Japan, 2-21-1 Osawa, Mitaka, Tokyo 181-8588, Japan}}
\newcommand{\IfA}{\affiliation{Institute for Astronomy, University of Hawai`i, 2680 Woodlawn Drive, Honolulu, HI 96822-1839, USA}}
\newcommand{\UHhilo}{\affiliation{Institute for Astronomy, University of Hawai`i, 640 N.~Aʻohoku Pl., Hilo, HI 96720, USA}}
\newcommand{\UCSC}{\affiliation{Department of Astronomy and Astrophysics, University of California, Santa Cruz, CA 95064-1077, USA}}
\newcommand{\Purdue}{\affiliation{Department of Physics and Astronomy, Purdue University, 525 Northwestern Avenue, West Lafayette, IN 47907-2036, USA}}
\newcommand{\Princeton}{\affiliation{Department of Astrophysical Sciences, Princeton University, 4 Ivy Lane, Princeton, NJ 08540-7219, USA}}
\newcommand{\Moore}{\affiliation{Gordon and Betty Moore Foundation, 1661 Page Mill Road, Palo Alto, CA 94304-1209, USA}}
\newcommand{\Durham}{\affiliation{Department of Physics, Durham University, South Road, Durham, DH1 3LE, UK}}
\newcommand{\JHU}{\affiliation{Department of Physics and Astronomy, The Johns Hopkins University, 3400 North Charles Street, Baltimore, MD 21218, USA}}
\newcommand{\Toronto}{\affiliation{David A.\ Dunlap Department of Astronomy and Astrophysics, University of Toronto,\\ 50 St.\ George Street, Toronto, Ontario, M5S 3H4 Canada}}
\newcommand{\Duke}{\affiliation{Department of Physics, Duke University, Campus Box 90305, Durham, NC 27708, USA}}
\newcommand{\NCU}{\affiliation{Graduate Institute of Astronomy, National Central University, 300 Jhongda Road, 32001 Jhongli, Taiwan}}
\newcommand{\Columbia}{\affiliation{Department of Physics and Columbia Astrophysics Laboratory, Columbia University, Pupin Hall, New York, NY 10027, USA}}
\newcommand{\Flatiron}{\affiliation{Center for Computational Astrophysics, Flatiron Institute, 162 5th Avenue, New York, NY 10010-5902, USA}}
\newcommand{\CIERA}{\affiliation{Center for Interdisciplinary Exploration and Research in Astrophysics and Department of Physics and Astronomy, \\Northwestern University, 1800 Sherman Avenue, 8th Floor, Evanston, IL 60201, USA}}
\newcommand{\GeminiNorth}{\affiliation{Gemini Observatory, 670 North A`ohoku Place, Hilo, HI 96720-2700, USA}}
\newcommand{\SNU}{\affiliation{Department of Physics and Astronomy, Seoul National University, Gwanak-ro 1, Gwanak-gu, Seoul, 08826, South Korea; scyoon\@snu.ac.kr}}

\newcommand{\GeminiNOIRLab}{\affiliation{Gemini Observatory/NSF's National Optical-Infrared Astronomy Research Laboratory, 670 N. Aohoku Place, Hilo, HI, 96720, USA; tom.geballe\@noirlab.edu}}

\newcommand{\Keck}{\affiliation{W.~M.~Keck Observatory, 65-1120 M\=amalahoa Highway, Kamuela, HI 96743-8431, USA}}
\newcommand{\UW}{\affiliation{Department of Astronomy, University of Washington, 3910 15th Avenue NE, Seattle, WA 98195-0002, USA}}
\newcommand{\catalyst}{\altaffiliation{LSST-DA Catalyst Fellow}}
\newcommand{\USask}{\affiliation{Department of Physics \& Engineering Physics, University of Saskatchewan, 116 Science Place, Saskatoon, SK S7N 5E2, Canada}}
\newcommand{\Thacher}{\affiliation{Thacher School, 5025 Thacher Road, Ojai, CA 93023-8304, USA}}
\newcommand{\Rutgers}{\affiliation{Department of Physics and Astronomy, Rutgers, the State University of New Jersey,\\136 Frelinghuysen Road, Piscataway, NJ 08854-8019, USA}}
\newcommand{\FSU}{\affiliation{Department of Physics, Florida State University, 77 Chieftan Way, Tallahassee, FL 32306-4350, USA}}
\newcommand{\Melbourne}{\affiliation{School of Physics, The University of Melbourne, Parkville, VIC 3010, Australia}}
\newcommand{\ASTROthreeD}{\affiliation{ARC Centre of Excellence for All Sky Astrophysics in 3 Dimensions (ASTRO 3D)}}
\newcommand{\Stromlo}{\affiliation{Mt.\ Stromlo Observatory, The Research School of Astronomy and Astrophysics, Australian National University, ACT 2601, Australia}}
\newcommand{\NCPAS}{\affiliation{National Centre for the Public Awareness of Science, Australian National University, Canberra, ACT 2611, Australia}}
\newcommand{\TAMU}{\affiliation{Department of Physics and Astronomy, Texas A\&M University, 4242 TAMU, College Station, TX 77843, USA}}
\newcommand{\Mitchell}{\affiliation{George P.\ and Cynthia Woods Mitchell Institute for Fundamental Physics \& Astronomy, College Station, TX 77843, USA}}
\newcommand{\ESO}{\affiliation{European Southern Observatory, Alonso de C\'ordova 3107, Casilla 19, Santiago, Chile}}
\newcommand{\ICE}{\affiliation{Institute of Space Sciences (ICE, CSIC), Campus UAB, Carrer
de Can Magrans, s/n, E-08193 Barcelona, Spain}}
\newcommand{\IEEC}{\affiliation{Institut d'Estudis Espacials de Catalunya (IEEC), Edifici RDIT, Campus UPC, 08860 Castelldefels (Barcelona), Spain}}
\newcommand{\Warwick}{\affiliation{Department of Physics, University of Warwick, Gibbet Hill Road, Coventry CV4 7AL, UK}}
\newcommand{\Macquarie}{\affiliation{School of Mathematical and Physical Sciences, Macquarie University, NSW 2109, Australia}}
\newcommand{\AAARC}{\affiliation{Astronomy, Astrophysics and Astrophotonics Research Centre, Macquarie University, Sydney, NSW 2109, Australia}}
\newcommand{\Capodimonte}{\affiliation{INAF - Capodimonte Astronomical Observatory, Salita Moiariello 16, I-80131 Napoli, Italy}}
\newcommand{\INFNNapoli}{\affiliation{INFN - Napoli, Strada Comunale Cinthia, I-80126 Napoli, Italy}}
\newcommand{\ICRANet}{\affiliation{ICRANet, Piazza della Repubblica 10, I-65122 Pescara, Italy}}
\newcommand{\MSU}{\affiliation{Center for Data Intensive and Time Domain Astronomy, Department of Physics and Astronomy,\\Michigan State University, East Lansing, MI 48824, USA}}
\newcommand{\SETI}{\affiliation{SETI Institute,
339 Bernardo Ave, Suite 200, Mountain View, CA 94043, USA; jrho\@seti.org}}
\newcommand{\IAIFI}{\affiliation{The NSF AI Institute for Artificial Intelligence and Fundamental Interactions}}
\newcommand{\ANUC}{\affiliation{Department of Astronomy, AlbaNova University Center, Stockholm University, SE-10691 Stockholm, Sweden}}

\newcommand{\Konkoly}{\affiliation{Konkoly Observatory,  CSFK, Konkoly-Thege M. \'ut 15-17, Budapest, 1121, Hungary}}
\newcommand{\ELTE}{\affiliation{ELTE E\"otv\"os Lor\'and University, Institute of Physics, P\'azm\'any P\'eter s\'et\'any 1/A, Budapest, 1117 Hungary}}
\newcommand{\SZTE}{\affiliation{Department of Experimental Physics, University of Szeged, D\'om t\'er 9, Szeged, 6720, Hungary}}
\newcommand{\IdAlta}{\affiliation{Instituto de Alta Investigaci\'on, Sede Esmeralda, Universidad de Tarapac\'a, Av. Luis Emilio Recabarren 2477, Iquique, Chile}}
\newcommand{\Kavli}{\affiliation{Kavli Institute for Cosmological Physics, University of Chicago, Chicago, IL 60637, USA}}
\newcommand{\UofChicago}{\affiliation{Department of Astronomy and Astrophysics, University of Chicago, Chicago, IL 60637, USA}}
\newcommand{\Fermi}{\affiliation{Fermi National Accelerator Laboratory, P.O.\ Box 500, Batavia, IL 60510, USA}}
\newcommand{\Dartmouth}{\affiliation{Department of Physics and Astronomy, Dartmouth College, Hanover, NH 03755, USA}}
\newcommand{\Surrey}{\affiliation{Department of Physics, University of Surrey, Guildford GU2 7XH, UK}}
\newcommand{\NU}{\affiliation{Center for Interdisciplinary Exploration and Research in Astrophysics (CIERA), Northwestern University, Evanston, IL 60208, USA}}
\newcommand{\itagaki}{\affiliation{Itagaki Astronomical Observatory, Yamagata 990-2492, Japan}}
\newcommand{\UdChile}{\affiliation{Departamento de Astronomia, Universidad de Chile, Camino El Observatorio 1515, Las Condes, Santiago, Chile}}
\newcommand{\UVA}{\affiliation{Department of Astronomy, University of Virginia, Charlottesville, VA 22904, USA}}
\newcommand{\UCSD}{\affiliation{Department of Astronomy \& Astrophysics, University of California, San Diego, 9500 Gilman Drive, MC 0424, La Jolla, CA 92093-0424, USA}}

\newcommand{\MIT}{\affiliation{Department of Physics and Kavli Institute for Astrophysics and Space Research, Massachusetts Institute of Technology, 77 Massachusetts Avenue, Cambridge, MA 02139, USA}}

\newcommand{\TAPIR}{\affiliation{TAPIR, Mailcode 350-17, California Institute of Technology, Pasadena, CA 91125, USA}}
\newcommand{\RESCEU}{\affiliation{Research Center for the Early Universe (RESCEU), School of Science, The University of Tokyo,  Bunkyo-ku, Tokyo 113-0033, Japan}}
\newcommand{\UIUC}{\affiliation{Department of Astronomy, University of Illinois at Urbana-Champaign, 1002 W. Green St., IL 61801, USA}}
\newcommand{\Copenhagen}{\affiliation{DARK, Niels Bohr Institute, University of Copenhagen, Jagtvej 155, 2200 Copenhagen, Denmark}}
\begin{CJK*}{UTF8}{gbsn}

\author[0000-0002-7937-6371]{Yize Dong (董一泽)}
\CfA\IAIFI
\email[show]{yize.dong@cfa.harvard.edu}

\author[0000-0002-5814-4061]{V. Ashley Villar}
\CfA
\IAIFI
\email{ashleyvillar@cfa.harvard.edu} 

\author[0000-0002-2028-9329]{Anya Nugent}
\CfA
\email{anya.nugent@cfa.harvard.edu}

\author[0000-0002-0832-2974]{Griffin Hosseinzadeh}
\UCSD
\email{ghosseinzadeh@ucsd.edu}

\author[0000-0002-2445-5275]{Ryan~J.~Foley}
\UCSC
\email{foley@ucsc.edu}

\author[0000-0002-8526-3963]{Christa~Gall}
\Copenhagen
\email{christa.gall@nbi.ku.dk}

\author[0000-0003-0648-2402]{Monica Gallegos-Garcia}
\CfA
\email{monica.gallegos_garcia@cfa.harvard.edu}

\author[0000-0003-4175-4960]{Conor Ransome}
\UA
\email{cransome@arizona.edu}

\author[0000-0002-9158-750X]{Aidan~Sedgewick}
\Copenhagen
\email{aidan.sedgewick@nbi.ku.dk}

\author[0000-0002-6347-3089]{Daichi Tsuna} 
\TAPIR\RESCEU\CfA
\email{daichi.tsuna@cfa.harvard.edu}

\author[0000-0001-8818-0795]{Stefano Valenti}
\UCD
\email{valenti@ucdavis.edu}

\author[0009-0000-7177-9697]{Henna Abunemeh}
\UIUC
\email{habun2@illinois.edu}

\author[0000-0002-1895-6639]{Moira Andrews}
\LCO\UCSB
\email{mandrews@lco.global}

\author[0000-0002-4449-9152]{Katie~Auchettl}
\UCSC\Melbourne
\email{katie.auchettl@unimelb.edu.au}

\author[0000-0002-4924-444X]{K.\ Azalee Bostroem}
\altaffiliation{LSSTC Catalyst Fellow}
\UA
\email{bostroem@arizona.edu} 

\author[0000-0003-4263-2228]{David~A.~Coulter}
\JHU\STScI
\email{dcoulter@stsci.edu}

\author[0000-0001-5486-2747]{Thomas de Boer}
\IfA
\email{tdeboer@hawaii.edu}

\author{Kaylee de~Soto}
\CfA
\email{kaylee.de_soto@cfa.harvard.edu}

\author[0000-0002-6886-269X]{Diego~A.~Farias}
\Copenhagen
\email{diego.farias@nbi.ku.dk}

\author[0000-0003-4914-5625]{Joseph Farah}
\LCO 
\UCSB
\email{jfarah@lco.global}

\author[0000-0002-7197-9004]{Danielle Frostig}
\CfA
\email{danielle.frostig@cfa.harvard.edu}

\author[0000-0003-1015-5367]{Hua~Gao}
\IfA
\email{hgao@hawaii.edu}

\author[0000-0003-4906-8447]{Alex Gagliano}
\IAIFI
\CfA
\MIT
\email{gaglian2@mit.edu}

\author[0000-0003-2744-4755]{Emily Hoang}
\UCD
\email{emthoang@ucdavis.edu}

\author[0000-0003-4253-656X]{D.\ Andrew Howell}
\LCO
\UCSB
\email{ahowell@lco.global}

\author[0000-0003-3953-9532]{Willem~B.~Hoogendam}
\altaffiliation{NSF GRFP Fellow}
\IfA
\email{willemh@hawaii.edu}

\author[0000-0003-1059-9603]{Mark~E.~Huber}
\IfA
\email{mehuber7@hawaii.edu}

\author[0000-0002-6230-0151]{David~O.~Jones}
\UHhilo
\email{dojones@hawaii.edu}

\author[0000-0002-7272-5129]{Chien-Cheng~Lin}
\IfA
\email{cclin33@hawaii.edu}

\author[0000-0001-9589-3793]{Michael Lundquist}
\Keck
\email{lund0946@gmail.com}

\author[0000-0001-5807-7893]{Curtis McCully}
\LCO
\email{curtismccully@gmail.com}

\author[0009-0008-9693-4348]{Darshana Mehta}
\UCD
\email{ddmehta@ucdavis.edu}

\author[0000-0001-6806-0673]{Anthony~L.~Piro}
\Carnegie
\email{piro@carnegiescience.edu}

\author[0000-0002-7352-7845]{Aravind P. Ravi}
\UCD
\email{apazhayathravi@ucdavis.edu}

\author[0000-0002-7015-3446]{Nicol\'as Meza Retamal}
\UCD
\email{nemezare@ucdavis.edu}

\author[0000-0002-7559-315X]{C{\' e}sar Rojas-Bravo}
\NAOC\UAS
\email{cesar.rojasbravo@ucas.ac.cn}

\author[0000-0002-0840-6940]{S.~Karthik~Yadavalli}
\CfA
\email{karthik.yadavalli@cfa.harvard.edu}

\author[0000-0001-5233-6989]{Qinan~Wang}
\MIT
\email{qnwang@mit.edu}





\begin{abstract}
In this paper, we first present observations of SN~2024acyl, a normal Type Ibn supernova with a large projected offset ($\sim$35~kpc) from its host galaxy. The low star-formation rate measured at the explosion site raises the possibility that the progenitor of SN~2024acyl may not have been a massive star.
We then examine, more broadly, the spectral diversity of Type Ibn supernovae around 20--35 days after peak brightness and identify two distinct groups: Group I, which shows bluer rest-frame optical color and narrower He~I emission lines; and Group II, which shows redder rest-frame optical color and broader He~I lines. Group~I also tends to show higher peak luminosities. The diversity we identify appears to be closely connected to the diversity observed around peak and to persist into late phases ($>80$ days after peak). Given its redder color and broader He~I lines, we classify SN~2024acyl as belonging to Group II. Based on the current dataset, we find no clear connection between this spectral diversity and either the host environments of Type Ibn SNe or their pre-explosion activity. The observed diversity in Type Ibn SNe likely reflects differences in circumstellar material properties and/or explosion energetics. These differences could result from a range of progenitor properties, such as different helium star mass, orbital period and companion type if they are in binary systems, and may indicate fundamentally diverse progenitors. Whether a continuous distribution exists between the two groups remains to be determined and will require further data to explore.

\end{abstract}





\section{Introduction} 

\label{sec:intro}
A growing number of supernovae (SNe) show observational signatures of strong interaction between the SN ejecta and dense circumstellar materials (CSM). Such interaction can serve as a major source of radiative energy and produce bright transients \citep[e.g.,][]{Smith2017hsn..book..403S,Modjaz2019NatAs...3..717M}. The presence of dense CSM implies that the progenitor stars experienced episodes of intense mass loss prior to explosion, although the underlying mass-loss mechanisms and the nature of the progenitor stars remain uncertain. Studying these interacting SNe offers valuable insights into the final stages of stellar evolution. 

Type Ibn SNe \corr{are} a subclass of interacting SNe, characterized by the presence of narrow helium (He) lines in their spectra \citep{Foley2007ApJ...657L.105F,Pastorello2007Natur.447..829P,Pastorello2008MNRAS.389..113P}. \corr{These features indicate} interaction with He-rich CSM. SNe Ibn typically show fast-evolving light curves, while the peak brightness, light curve decline slope, and color evolution can be heterogeneous \citep{Hosseinzadeh2017ApJ...836..158H, Farias2025inprep}. Type Ibn SNe also show spectral diversity at early times: some events display narrow P-Cygni lines, while others show broader He emission lines, suggesting a range of CSM configurations or even multiple progenitor channels \citep{Hosseinzadeh2017ApJ...836..158H}.


Light curve modeling of Type Ibn SNe found that the He-rich CSM has to be dense and located close to the progenitor in order to explain the light curve evolution \citep{Wang2020ApJ...900...83W,Gangopadhyay2020ApJ...889..170G,Wang2021ApJ...917...97W,Pellegrino2022ApJ...926..125P, Ben-Ami2023ApJ...946...30B, Dong2024ApJ...977..254D,Wang2025A&A...700A.156W,Gangopadhyay2025MNRAS.tmp.1456G,Baer-Way2025arXiv250907080B}, pointing toward enhanced mass loss from He star progenitors shortly before explosion.
This has motivated searches for pre-explosion \corr{activity} in SNe Ibn. Thus far, confirmed precursors have been observed in only three cases: SN~2006jc \citep{Pastorello2007Natur.447..829P}, SN~2019uo \citep{Strotjohann2021ApJ...907...99S}, and SN~2023fyq \citep{Dong2024ApJ...977..254D,Brennan2024A&A...684L..18B}. SN~2006jc and SN~2019uo had brief outbursts 1--2~years before explosion, while SN~2023fyq showed long-lasting precursor emission for up to three years, \corr{attributed} to binary interactions \citep{Dong2024ApJ...977..254D,Tsuna2024OJAp....7E..82T}. Whether all SNe Ibn have detectable precursors remains unclear, but the diversity observed in these cases potentially implies multiple mechanisms for pre-explosion mass loss.



Radiative transfer spectral simulations suggest that the progenitors of SNe Ibn are likely low-mass He stars  \citep{Dessart2022A&A...658A.130D}. These models require solar metallicity to reproduce the strong blue continuum observed in some events. By applying these models to SN~2020nxt, \cite{Wang2024MNRAS.530.3906W} argue that at least some Type Ibn SNe likely arise from binary He-rich stars and thus not all SNe Ibn arise from single massive Wolf-Rayet stars.

Host-environment studies provide additional clues about the progenitor systems of Type Ibn SNe. Although most of SNe Ibn are found in star-forming spiral galaxies \citep{Pastorello2015MNRAS.449.1954P}, a few exceptions challenge the massive star scenario.
PS1-12sk exploded in an elliptical galaxy \citep{Sanders2013ApJ...769...39S} with a low star formation rate (SFR, \citealt{Hosseinzadeh2019ApJ...871L...9H}), and SN~2023tsz occurred in a low-mass galaxy with very low metallicity and low SFR \citep{Warwick2025MNRAS.536.3588W}.
Such environments are atypical for core-collapse SNe, which are generally linked to massive-star explosions. These cases therefore raise the question of whether all SNe Ibn originate from massive stars.





In this paper, we present the normal Type Ibn SN~2024acyl, located in a region of low SFR at a projected distance of approximately 35~kpc from its host galaxy. In addition, we identify spectral diversity among Type Ibn SNe at $\sim$20--35 days after peak light, which appears to be closely connected to the early-time diversity reported by \cite{Hosseinzadeh2017ApJ...836..158H}, and we introduce a more quantitative framework for characterizing this diversity. The paper is organized as follows: the observations of SN~2024acyl are presented in Section \ref{sec:observations}; the photometric and
spectroscopic evolution are described in Section \ref{sec:phot_evol} and \ref{sec:spec_evol}. The host environment of SN~2024acyl is discussed in Section \ref{sec:host_env}. In Section \ref{sec:spec_diversity}, we examine the spectral diversity of Type Ibn SNe around 20-35 days after peak and explore its implications. We summarize the main results in Section~\ref{sec:conclusions}.

\section{Observations} \label{sec:observations}
\begin{figure*}
\includegraphics[width=0.667\linewidth]{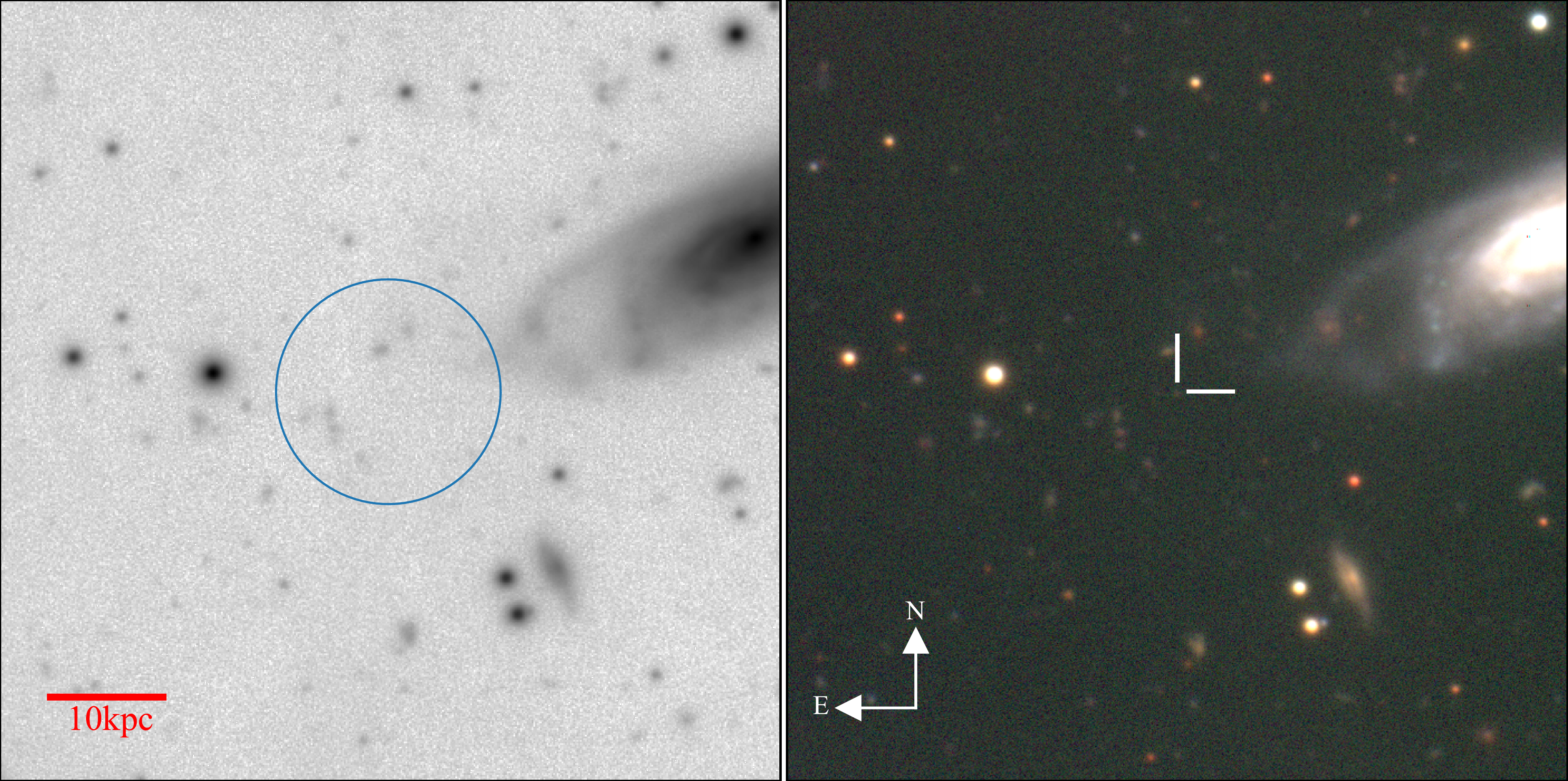}
\includegraphics[width=0.333\linewidth]{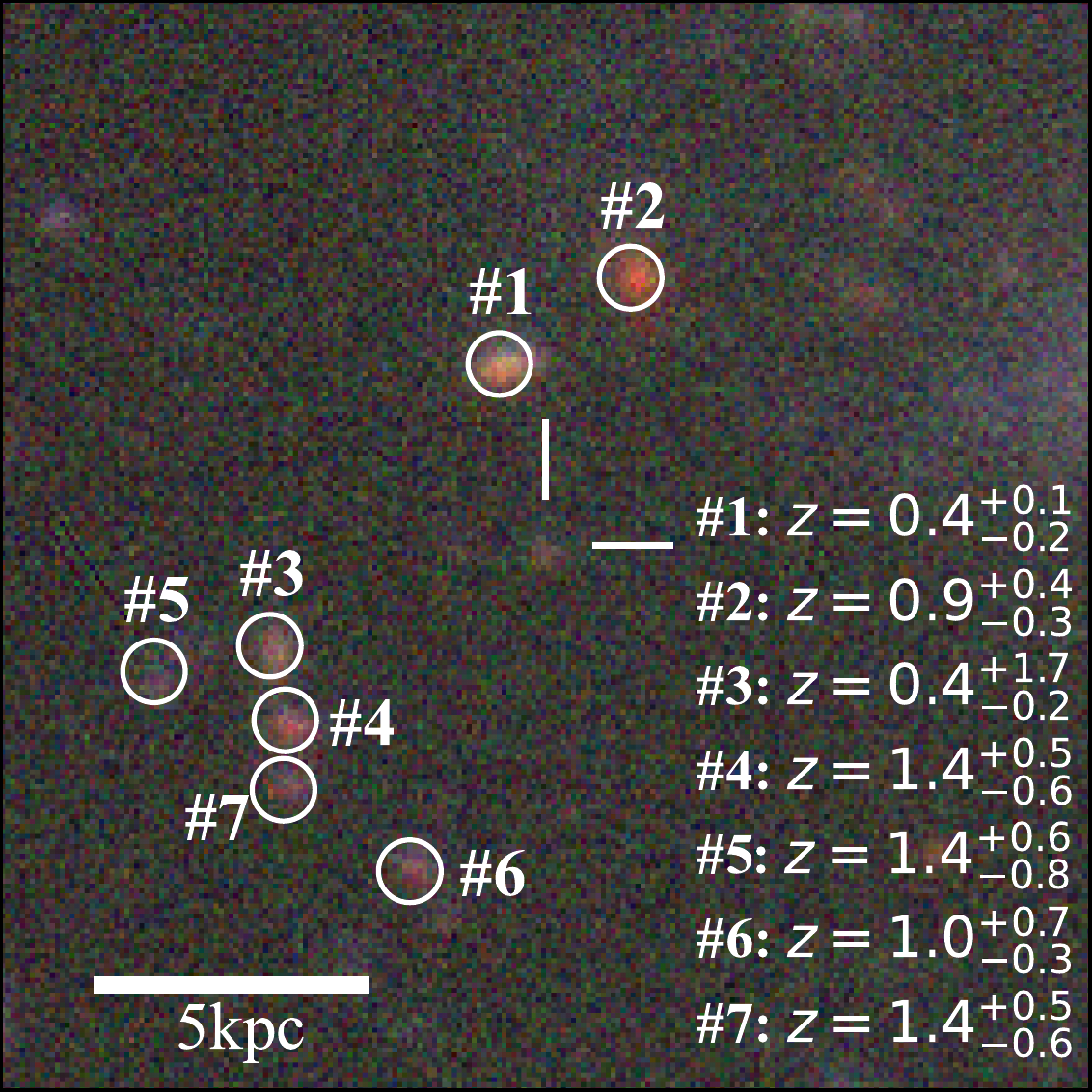}
\caption{Left: Stacked $w$-band image from PS1 and PS2 taken before the SN explosion. There is no source detected at the SN position down to 24.5~mag. The blue circle, centered on the SN location, has a radius of 10 kpc and is shown for visual reference. Middle: Composite $gri$ image of SN~2024acyl obtained with the Gemini-North Observatory on 2025 February 22. The position of SN~2024acyl is indicated by white tick markers. Right: A zoom-in on the region around SN~2024acyl. Seven extended sources near the SN are labeled, with their estimated redshifts indicated.
\label{fig:sn_image}}
\end{figure*}

SN~2024acyl was discovered on 2024 December 1 by the Asteroid Terrestrial-Impact Last Alert System (ATLAS, \citealt{Tonry2018}) survey at RA(2000) $=$ 02\textsuperscript{h}46\textsuperscript{m}05\fs326, Dec(2000) $= +28\degr 01\arcmin 17\farcs 91$ \citep{sn2024acyl_discovery}, located in CGCG~505-052 (see Figure \ref{fig:sn_image}), with an offset of $\sim59\farcs6$ from the host galaxy center. On 2024 December 6, SN~2024acyl was classified as a Type Ibn \citep{sn2024acyl_classification}. 

The redshift of SN~2024acyl is measured to be $z=0.027^{+0.001}_{-0.001}$ from the early-time narrow He emission lines in the spectra. Adopting a cosmology with $H_{0}=67.4 ~\rm km\,s^{-1}\,Mpc^{-1}$ and $\Omega_{m}=0.315$ \citep{Planck2016A&A...594A..13P,Planck2020A&A...641A...6P}, this corresponds to a luminosity distance of $122^{+5}_{-5}$~Mpc, which will be used throughout our analysis.
The Milky Way line-of-sight reddening toward SN~2024acyl is $E(B-V)$ = 0.13 mag \citep{Schlafly2011ApJ...737..103S}. Given the large offset between the object and host galaxy, and the fact that there are no obvious narrow Na~I~D absorptions lines from the optical spectra, 
the host extinction is likely low.
Therefore, in our analysis, we only account for extinction from the Milky Way.

\begin{deluxetable}{cc}
\tablecaption{Basic properties of SN~2024acyl\label{tab:sn_properties}}
\tablewidth{0pt}
\tablehead{
}
\startdata
Host galaxy & CGCG~505-052\\
RA (J2000) & 02\textsuperscript{hr}46\textsuperscript{m}05\fs33\\
DEC (2000) & $+28\degr 01\arcmin 17\farcs 91$\\
Distance & $122^{+5}_{-5}$~Mpc\\
Distance modulus& $35.4_{-0.1}^{+0.1}$~mag \\
Redshift $z$ & $0.027^{+0.001}_{-0.001}$\\
$E(B-V)_{\rm MW}$ & $0.13_{-0.01}^{+0.01}$ mag$^\star$\\
Explosion epoch (JD) & $\rm JD~2,460,642.5\pm0.1$ (2024-11-28)\\
$o_{\rm max}$ (JD)$^{+}$ & $\rm JD~2,460,653.8\pm0.2$ (2024-12-09)\\
\enddata
\tablenotetext{\star}{\cite{Schlafly2011ApJ...737..103S}.}
\tablenotetext{$+$}{The peak epoch derived from the $o$-band light curve.
}
\end{deluxetable}

\begin{figure*}
\centering
\includegraphics[width=0.9\linewidth]{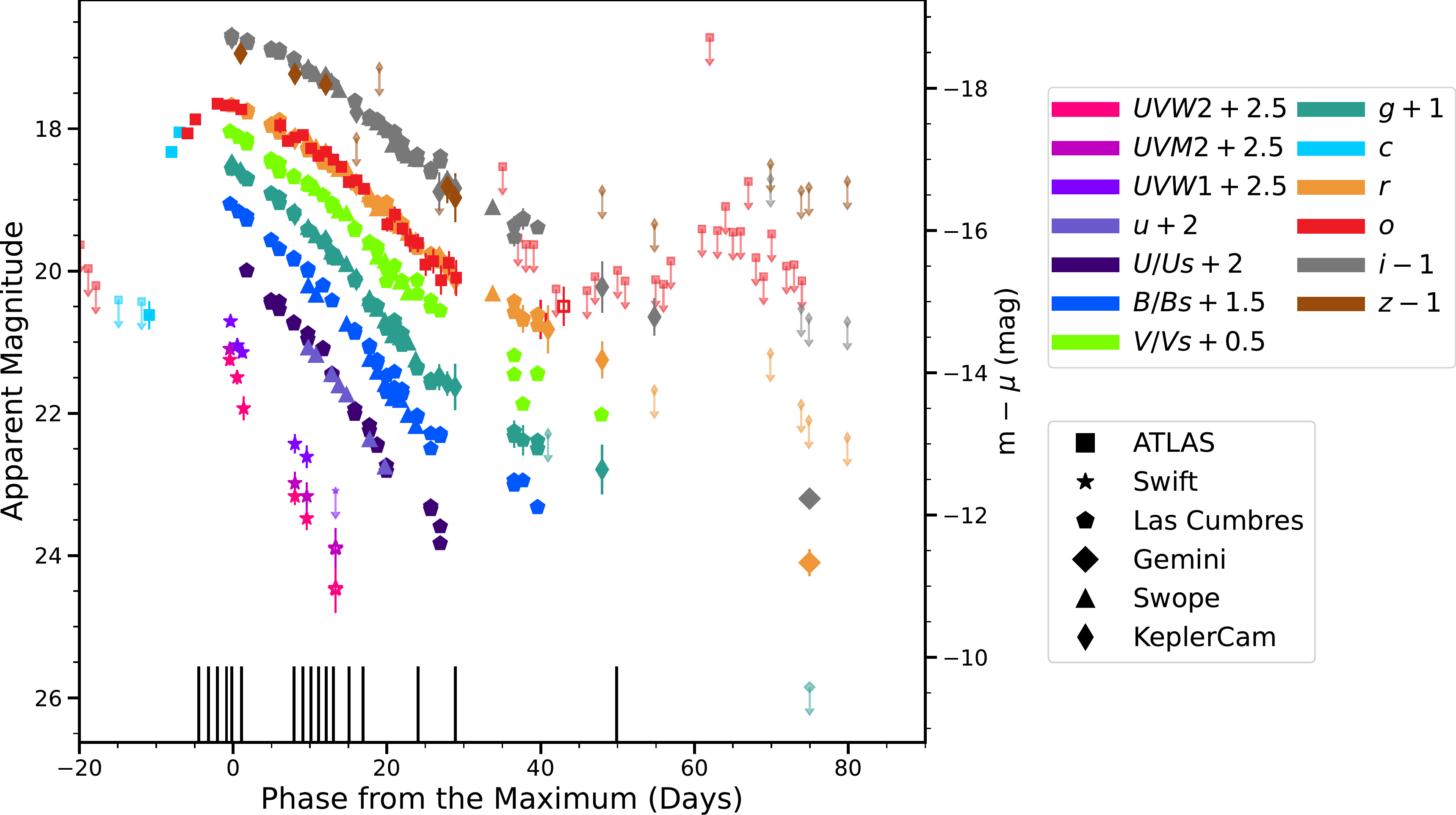}
\caption{Photometric evolution of SN~2024acyl with respect to the maximum light. Detections with S/N~$>4$ are shown as large filled symbols, those with $3 < \mathrm{S/N} \leq 4$ as hollow symbols, and non-detections with $\mathrm{S/N} \leq 3$ as smaller downward limits. Epochs of spectra are marked by black lines along the bottom axis.
\label{fig:lc_evol}}
\end{figure*}

\begin{figure*}
\centering
\includegraphics[width=0.9\linewidth]{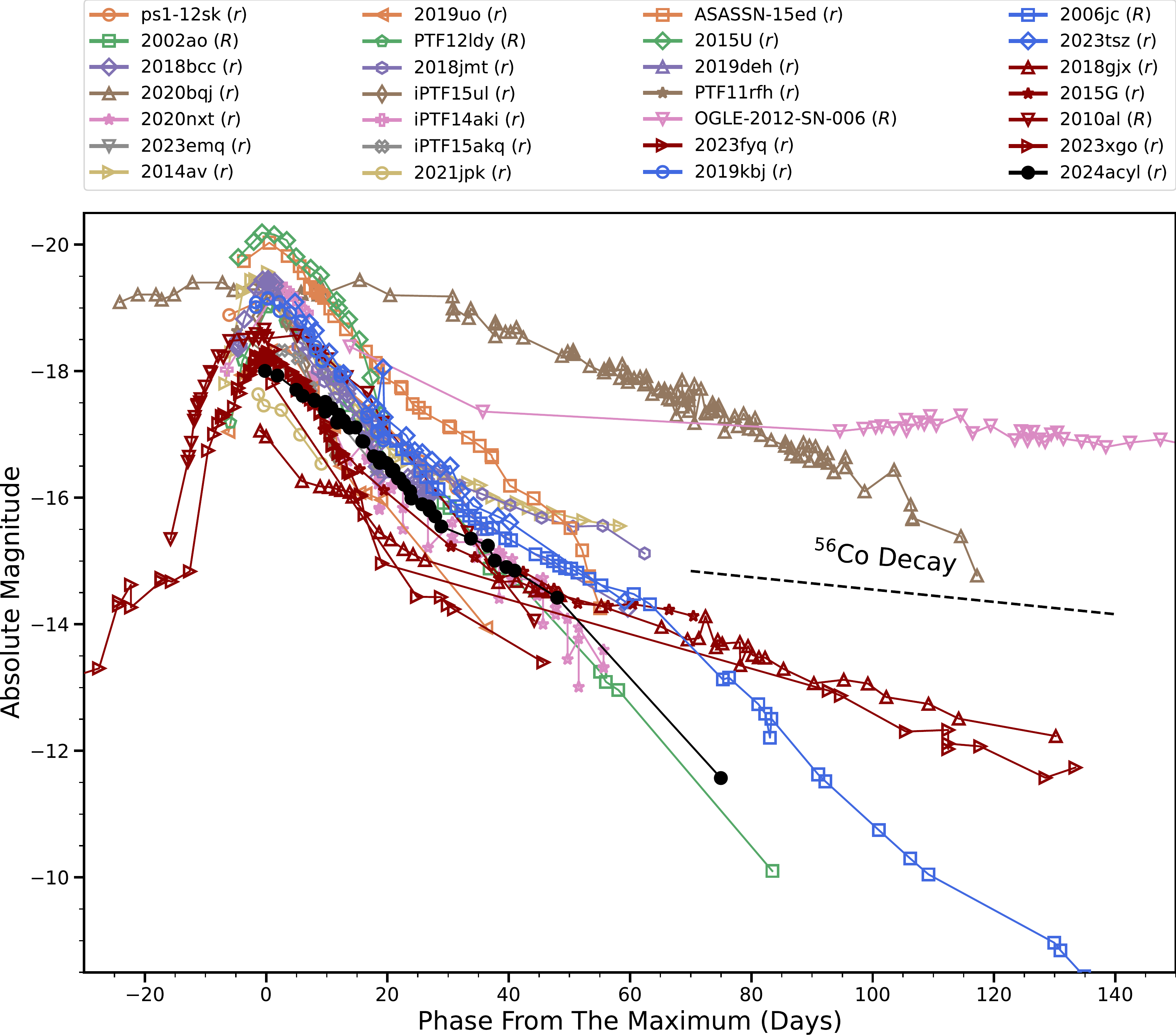}
\caption{$r/R$-band light curve comparison between SN~2024acyl and 27 Type Ibn SNe with well-sampled light curves. \corr{The $^{56}$Co decay rate is shown as a dashed line for comparison.}
The Type Ibn SNe used in this plot includes: 
PS1-12sk \citep{Sanders2013ApJ...769...39S},
SN~2002ao \citep{Pastorello2008MNRAS.389..113P},
SN~2018bcc \citep{Karamehmetoglu2021AA...649A.163K},
SN~2020bqj \citep{Kool2021AA...652A.136K},
SN~2020nxt \citep{Wang2024MNRAS.530.3906W}, 
SN~2023emq \citep{Pursiainen2023ApJ...959L..10P},
SN~2014av \citep{Pastorello2016MNRAS.456..853P},
SN~2019uo \citep{Gangopadhyay2020ApJ...889..170G}, 
PTF12ldy \citep{Hosseinzadeh2017ApJ...836..158H},
SN~2018jmt \citep{Wang2024A&A...691A.156W},
iPTF15ul \citep{Hosseinzadeh2017ApJ...836..158H},
iPTF14aki \citep{Hosseinzadeh2017ApJ...836..158H},
iPTF15akq \citep{Hosseinzadeh2017ApJ...836..158H},
SN~2021jpk \citep{Pellegrino2022ApJ...926..125P},
ASASSN-15ed \citep{Pastorello2015MNRAS.453.3649P}, 
SN~2015U \citep{Tsvetkov2015IBVS.6140....1T,Pastorello2015MNRAS.454.4293P,Hosseinzadeh2017ApJ...836..158H}, 
SN~2019deh \citep{Pellegrino2022ApJ...926..125P},
PTF11rfh \citep{Hosseinzadeh2017ApJ...836..158H},
OGLE-2012-SN-006 \citep{Pastorello2015MNRAS.449.1941P},
SN~2023fyq \citep{Dong2024ApJ...977..254D, Brennan2024A&A...684L..18B},
SN~2019kbj \citep{Ben-Ami2023ApJ...946...30B},
SN~2006jc \citep{Pastorello2007Natur.447..829P}, 
SN~2023tsz \citep{Warwick2025MNRAS.536.3588W},
SN~2018gjx \citep{Prentice2020MNRAS.499.1450P}, 
SN~2015G \citep{Shivvers2017MNRAS.471.4381S, Hosseinzadeh2017ApJ...836..158H}, 
and SN~2010al \citep{Pastorello2015MNRAS.449.1921P}, 
\label{fig:lc_comp}}
\end{figure*}

\subsection{Photometric Observations}
Following the discovery of SN~2024acyl, we initiated a multiband photometric follow-up campaign using the Las Cumbres Observatory Global Telescope Network \citep{Brown2013PASP..125.1031B} via the Global Supernova Project, the 1-meter Henrietta Swope telescope at Las Campanas Observatory, the KeplerCam instrument on the FLWO 1.2-m telescope at the Fred Lawrence Whipple Observatory, the Gemini Multi-Object Spectrograph \citep[GMOS;][]{Hook2004} on Gemini-North through proposal GN-2025A-Q-208, and the UVOT instrument \citep{Roming2005SSRv..120...95R} on the Neil Gehrels \textit{Swift} Observatory \citep{Gehrels2004}.

We obtained photometry from the ATLAS forced photometry server \citep{Smith2020PASP..132h5002S,Shingles2021}. We stacked the measurements using the method presented in \cite{Young2022}, to improve the signal-to-noise ratio and obtain deeper upper limits. 
\corr{
TESS observed the location of SN~2024acyl between JD~2,460,610.6--2,460,636.1, and the data were retrieved from the Mikulski Archive for Space Telescopes (MAST\footnote{https://archive.stsci.edu/}).
}In addition, we retrieved pre-explosion $w$- and $i$-band images of SN~2024acyl from the Pan-STARRS Survey for Transients \citep[PSST;][]{Huber2015IAUGA..2258303H}, spanning from 2015 October 24 to 2024 November 23.
The reduction process for the photometric data is presented
in Appendix \ref{append:data_reduc}.

\subsection{Spectrocopic Observations}
The spectroscopic observations of SN~2024acyl were taken with the FLOYDS spectrograph \citep{Brown2013PASP..125.1031B} on the 2m Faulkes Telescope North in Hawaii at the Las Cumbres Observatory via the Global Supernova Project, the Binospec instrument on the MMT \citep{Binospec} through a follow-up program on behalf of the Young Supernova Experiment, GMOS \citep{Hook2004} on Gemini-North \corr{through proposal GN-2025A-Q-208}, the Low-Resolution Imaging Spectrometer \citep[LRIS;][]{Oke1995} on the 10~m Keck~I telescope, MOSFIRE \citep{McLean10.1117/12.856715} on the Keck II telescope, and SpeX \citep{Rayner2003PASP..115..362R} on the NASA Infrared Telescope Facility (IRTF). \corr{A host spectrum of CGCG~505-052 was also obtained with the FAST Spectrograph \citep{Fabricant1998PASP..110...79F} on the Fred L. Whipple Observatory 1.5-meter Tillinghast telescope.}


In addition, we include three publicly available spectra from the Transient Name Server (TNS) in our analysis: two taken with EFOSC2 on the NTT on 2024 December 06 \citep{Santos2024TNSCR4787....1S} and 2024 December 09 \citep{Duarte2024TNSCR4837....1D}, and one obtained on 2024 December 04 using an ALPY 200 grism-slit spectrograph on a 0.3-meter aperture telescope \citep{Soubrouillard2024TNSCR4856....1S}.

We also present an unpublished spectrum of the Type Ibn SN~2023xgo, taken with Keck~I LRIS on 2023 December 12; a detailed analysis of this object has been presented in \cite{Gangopadhyay2025MNRAS.tmp.1456G}. In addition, a nebular-phase spectrum of Type Ibn SN~2023tsz (Vasylyev et al., in prep.), obtained with Keck~I LRIS on 2023 December 12, is included in our analysis.


A log of the spectroscopic observations is given in Table \ref{appendix:spec_table}.
The reduction process for the spectroscopic data is presented
in Appendix \ref{append:data_reduc}.

\begin{figure*}
\centering
\includegraphics[width=0.85\linewidth]{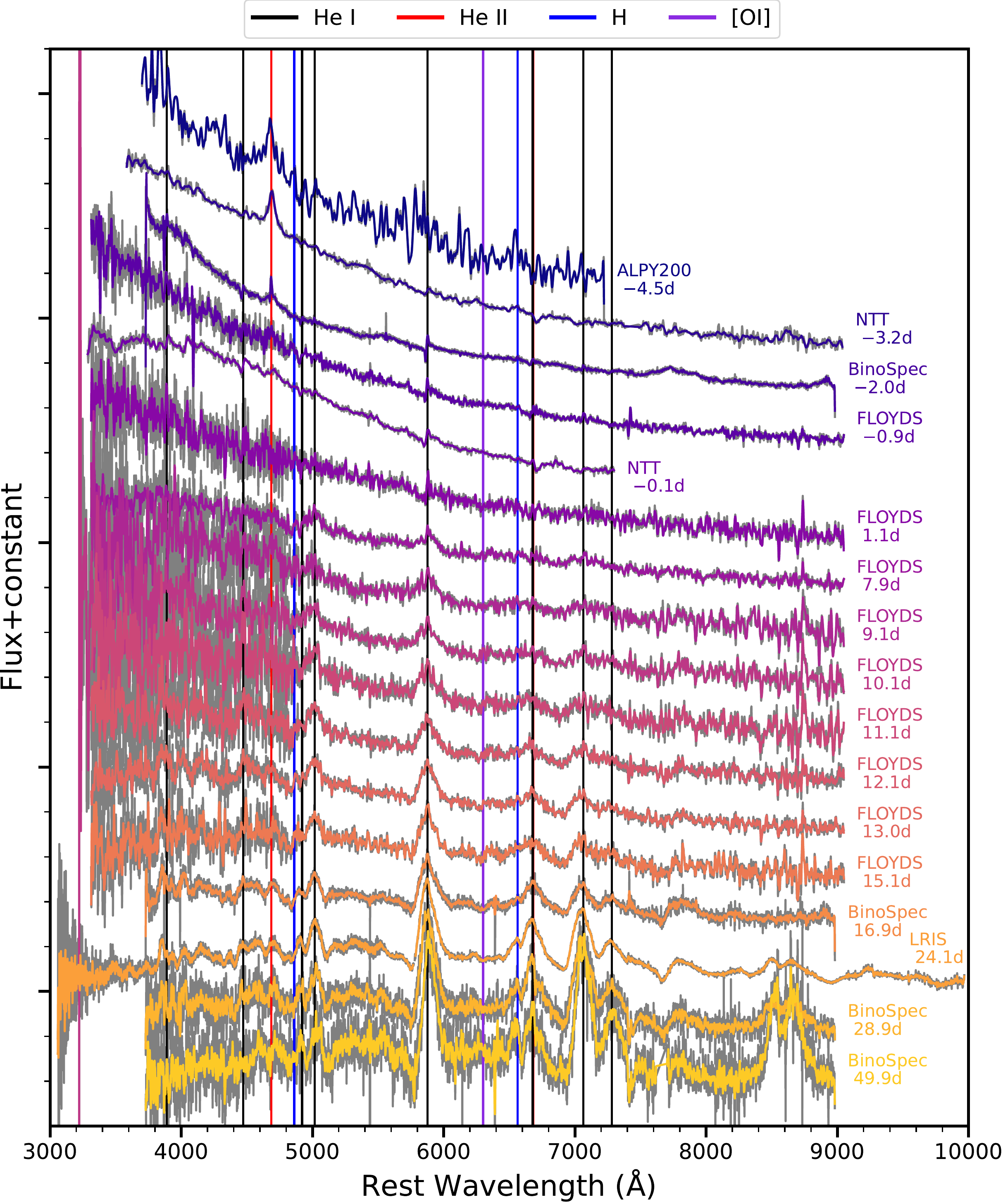}
\caption{Spectrocopic evolution of SN~2024acyl. The phase is measured from the $o$-band maximum. The spectra shown here have been smoothed with a second order Savitzky-Golay filter, and the gray background lines are the original spectra.
\label{fig:spec_evol}}
\end{figure*}

\section{Photometric Evolution} \label{sec:phot_evol}
In Figure \ref{fig:lc_evol}, we present the photometric evolution of SN~2024acyl. \corr{The $gri$-band photometry are in AB system, while the $UBV$-band photometry are in Vega system.} The SN was first detected by ATLAS in the $c$-band on JD~2,460,642.9. The last nondetection was on JD~2,460,641.9, with a 5$\sigma$ limit of 20.4~mag, constraining the explosion epoch to within this interval. As a sanity check, we fit the early \corr{pre-peak} $c$-band \corr{data points} in flux space with a power law in the form $f \propto(\frac{t-t_{0}}{1+z})^{n}$, where $t_{0}$ is the explosion epoch. \corr{This gives $t_{0}=\rm JD~2,460,642.5\pm0.1$ while $n=0.9\pm0.1$.} We measure the maximum of $o$-band light curve to be $\rm JD~2,460,653.8\pm0.2$ by fitting the light curve with a spline function and infer a rise time of $11.3\pm0.2$ days. We summarize the light curve measurements along with other basic properties in Table~\ref{tab:sn_properties}.

In Figure \ref{fig:lc_comp}, we compare the $r$-band light curve of SN~2024acyl with those of 27 Type Ibn SNe that have well-sampled light curves in the $r/R$ band. 
All the $R$-band magnitudes have been shifted to the AB magnitude system. The distance modulus \corr{as well as Galactic and host} extinction values for the objects in the comparison sample were obtained from the corresponding papers cited in the caption of Figure~\ref{fig:lc_comp}.
The light curve evolution of SN~2024acyl is broadly similar to other SNe Ibn. 
Previous studies show that the fast-evolving and luminous light curves of Type Ibn SNe generally cannot be explained with solely a $\rm ^{56}Ni$ decay model \citep{Ben-Ami2023ApJ...946...30B, Gangopadhyay2020ApJ...889..170G, Dong2024ApJ...977..254D}.
We therefore model the multiband light curves of SN~2024acyl with a combined CSM+$\rm ^{56}Ni$ decay model using MOSF{\sc i}T \citep{Guillochon2018ApJS..236....6G}. 
The $\rm ^{56}Ni$ decay and CSM interaction models are based on \cite{Arnett1982} and \cite{Chatzopoulos2013ApJ...773...76C}, respectively, and the details of the light curve fitting are described in Appendix \ref{append:lc_fit}. The best-fit model requires an ejecta mass of $\rm 0.70^{+0.20}_{-0.10}\ M_{\odot}$, a CSM mass of $\rm 0.36^{+0.04}_{-0.03}\ M_{\odot}$, and a $\rm ^{56}Ni$ mass of $\rm 0.03^{+0.01}_{-0.01}\ M_{\odot}$. These values are consistent with those found in previous studies for individual objects \citep{Pellegrino2022ApJ...926..125P,Ben-Ami2023ApJ...946...30B} \corr{and a sample study of Type Ibn SNe \citep{Farias2025inprep}}.

\corr{The location of SN~2024acyl was monitored by ATLAS and TESS roughly five years and one month prior to the explosion, respectively. No signs of pre-explosion variability were detected, with limiting magnitudes of $o > 20.4$ and $m_{\rm TESS} > 18.8$, ruling out precursor emission brighter than $\sim-15$~mag (ATLAS) and $\sim-17$~mag (TESS).}

\begin{figure}
\centering
\includegraphics[width=1.0\linewidth]{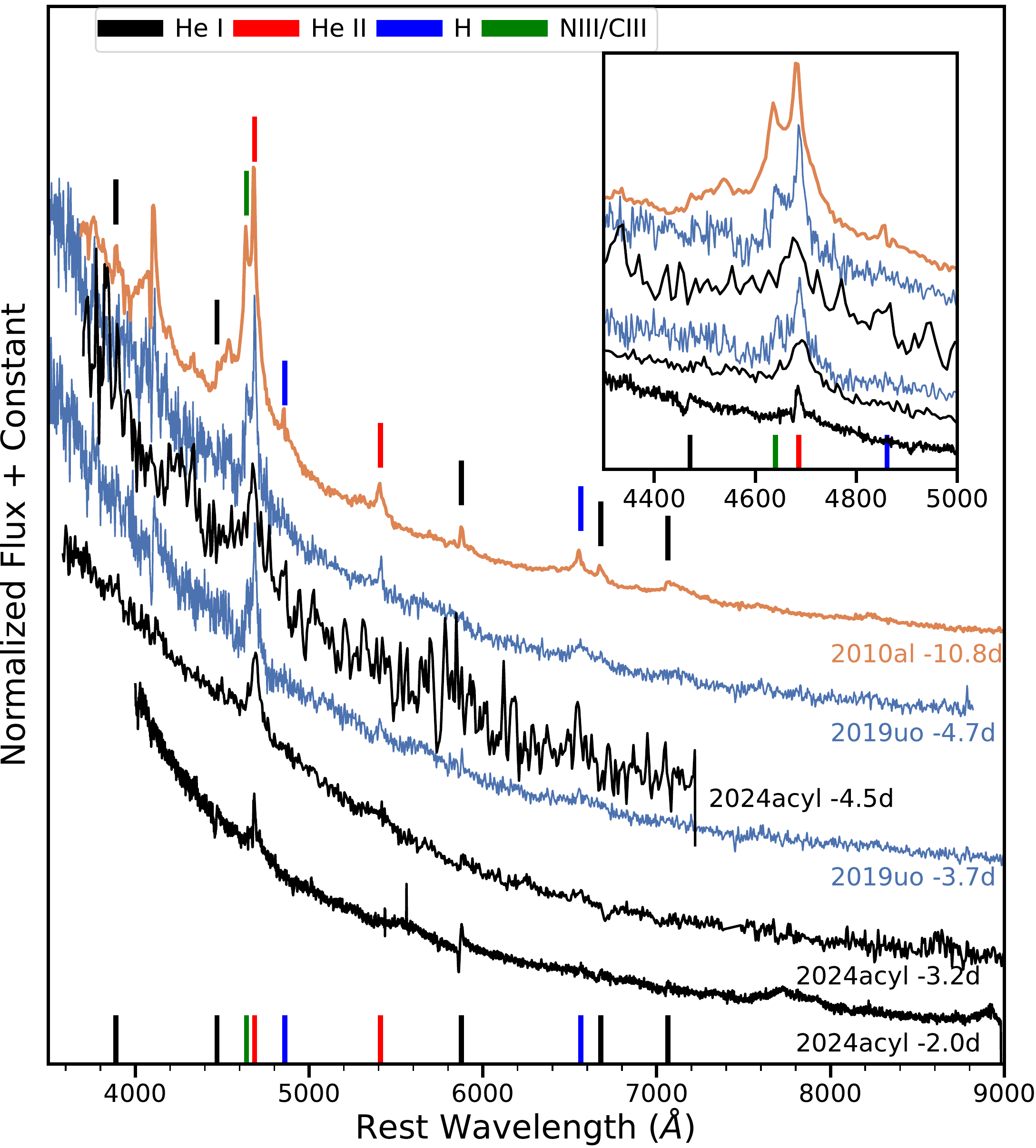}
\caption{
Spectral evolution of SN~2024acyl at -4.1 days, -2.8 days and -1.6 days compared to SN~2010al and SN~2019uo.
The inset shows the evolution of the flash ionization lines He~II $\lambda4686$ and N~III $\lambda4640$/C~III $\lambda4650$, which become weaker over time, while He~II $\lambda4686$ develops a P-Cygni profile at -2.0 days.
\label{fig:flash}}
\end{figure}

\section{Spectroscopic Evolution} \label{sec:spec_evol}
The spectral evolution of SN~2024acyl is shown in Figure \ref{fig:spec_evol}. Prior to peak brightness, SN~2024acyl shows a blue continuum with flash ionization lines of He~II $\lambda4686$, N~III $\lambda4640$/C~III $\lambda4650$ and other He~I features. In Figure \ref{fig:flash}, we compare the early spectra of SN~2024acyl with those of two other well-observed SNe Ibn that showed flash ionization features: SN~2010al and SN~2019uo. 
\corr{In SN~2024acyl, although the spectral resolution is low, the $\rm H\alpha$ line may be present at $-4.5$ days. At $-3.2$ and $-2.0$ days, a weak $\rm H\alpha$ feature is visible in the spectra. The presence of $\rm H\alpha$ is similar to that observed in SN~2010al and SN~2019uo \citep{Pastorello2015MNRAS.449.1921P,Gangopadhyay2020ApJ...889..170G}.}
At $-2.0$ days, He~I $\lambda5876$ and He~II $\lambda4686$ show narrow P-Cygni features (Figure \ref{fig:flash}), likely formed in the unshocked He-rich CSM. Such narrow P-Cygni line profiles have been observed in many Type Ibn SNe around peak \citep[e.g.,][]{Pastorello2015MNRAS.449.1921P,Hosseinzadeh2017ApJ...836..158H,Gangopadhyay2020ApJ...889..170G}.

Approximately one week after peak, broader He~I lines begin to develop in the spectra of SN~2024acyl. Starting around +24 days, an additional emission feature appears on the blue wing of He~I $\lambda6678$ which is likely $\rm H\alpha$. Hydrogen emission lines have been observed in post-peak spectra of other Type Ibn SNe \citep{Pastorello2008MNRAS.389..131P,Karamehmetoglu2021AA...649A.163K}, which are often interpreted as signatures of an outer hydrogen-rich CSM shell. \corr{For a more detailed post-peak spectral modeling of SN~2024acyl, we refer the reader to \cite{Cai2025arXiv251104337C}.}

\begin{figure}
\includegraphics[width=1.\linewidth]{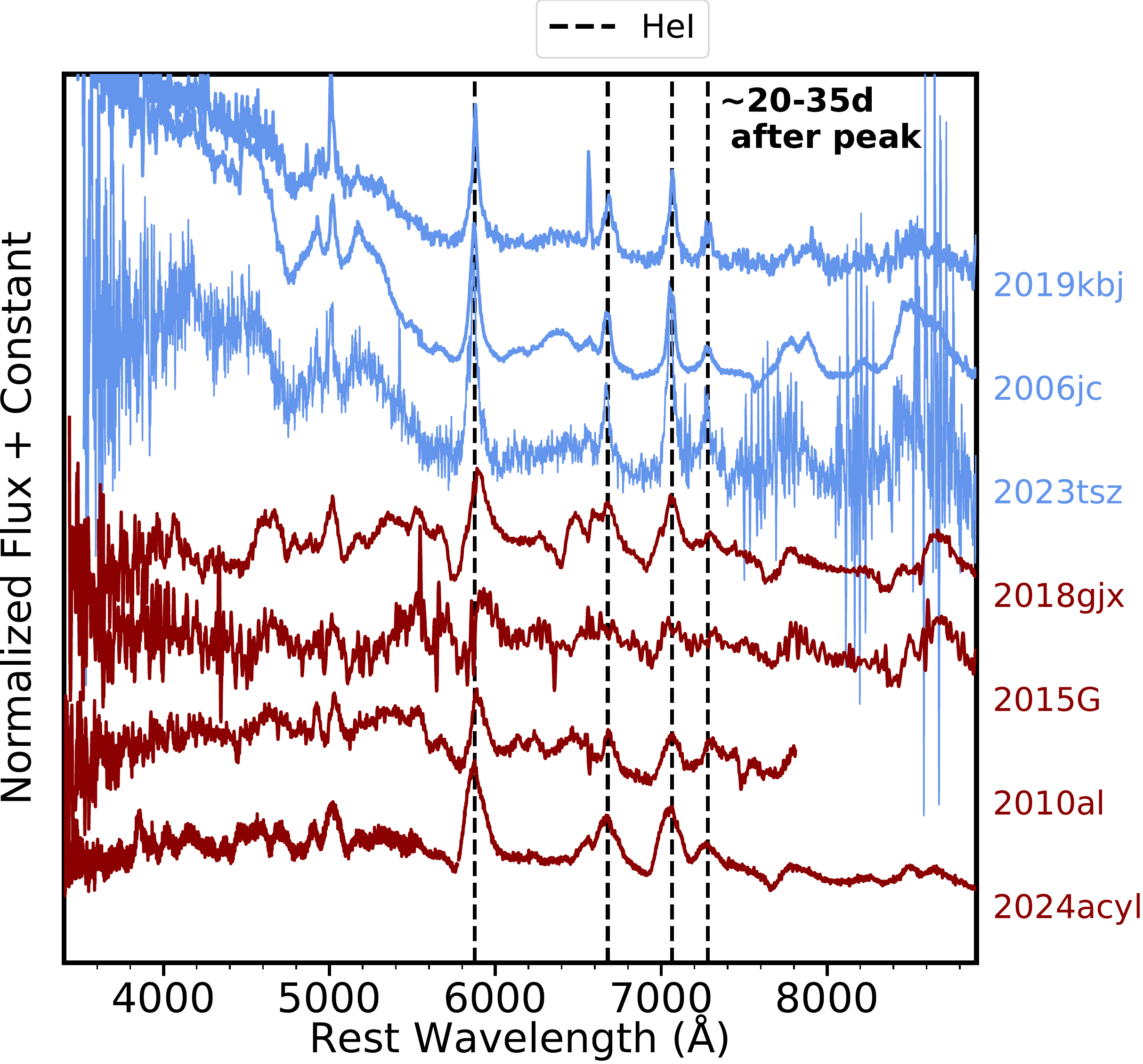}
\caption{Spectral comparison of SN~2024acyl at +24.1 days with other Type Ibn SNe at $\sim20-35$ days. \corr{Some SNe Ibn show a strong blue continuum blueward of $\sim$5500~$\rm \AA$, arising from the Fe~II forest, while others do not. Objects with strong blue continua are shown in blue (Group~I), and those with weaker continua are shown in red (Group~II).}
\label{fig:spec_comp}}
\end{figure}

We compare the spectrum of SN~2024acyl taken at +24.1 days with those of other Type Ibn SNe taken at $\sim$20--35 days after peak in Figure \ref{fig:spec_comp}. 
In this comparison sample, we include SN~2006jc and SN~2010al, the prototypes of the Type Ibn class, along with other well-observed Type Ibn SNe. SN~2006jc, SN~2019kbj, and SN~2023tsz show a strong blue continuum blueward of $\sim$5500~$\rm \AA$, which is often attributed to a forest of Fe~II emission lines \citep{Foley2007ApJ...657L.105F,Smith2008ApJ...680..568S,Pastorello2016MNRAS.456..853P,Dessart2022A&A...658A.130D}. In contrast, SN~2024acyl lacks this kind of blue emission, similar to SN~2010al, SN~2015G, and SN~2018gjx. 
In addition, the He~I emission lines in SN~2006jc, SN~2019kbj, and SN~2023tsz tend to be \textit{narrower} than those in SN~2024acyl, SN~2010al, SN~2015G, and SN~2018gjx.
This diversity will be discussed further in Section \ref{sec:spec_diversity}.

\begin{figure}
\includegraphics[width=1.\linewidth]{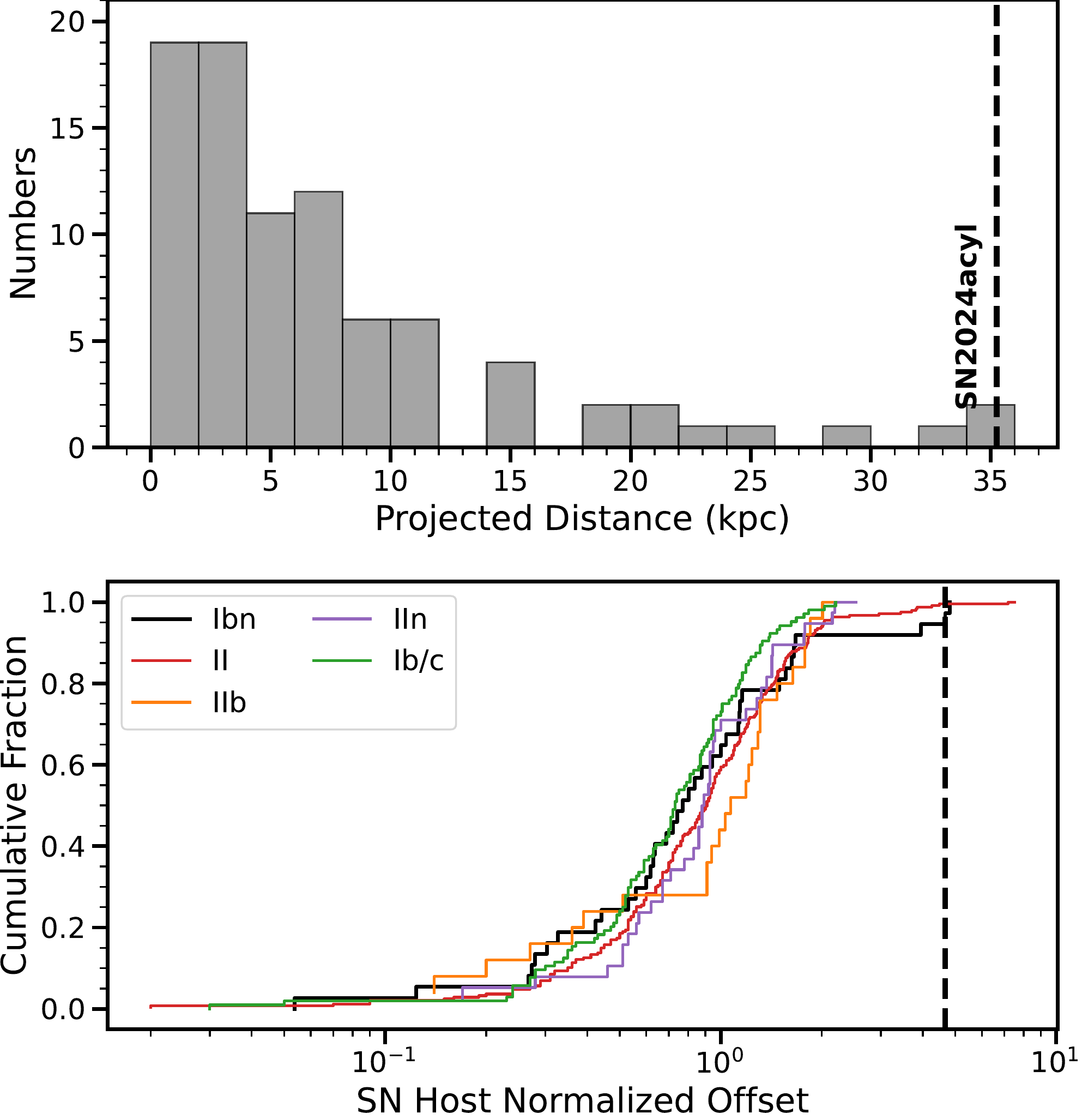}
\caption{Upper: The projected offset distribution of Type Ibn SNe. The black vertical line marks SN~2024acyl. Bottom: The host offset normalized by half-light radius of the host galaxies. The black vertical line marks SN~2024acyl. 
\label{fig:host_offset}}
\end{figure}

\section{Host Environments of SN~2024acyl} \label{sec:host_env}
The host environments of SNe can provide valuable insights into their progenitors and explosion mechanisms \citep[e.g.,][]{Hakobyan2012A&A...544A..81H,Galbany2018ApJ...855..107G}. \corr{The explosion site of SN~2024acyl is significantly offset from any obvious host galaxies (Figure~\ref{fig:sn_image}), motivating a detailed investigation of its host environment.}

\subsection{Global Host Environment}
\label{sec:global_host}

\corr{The redshift of SN~2024acyl ($z = 0.027$) is consistent with that of CGCG~505-052 \citep{Zaw2019ApJ...872..134Z}, making it highly likely that SN~2024acyl is associated with this galaxy. This host association is further supported by \texttt{Pr\"ost}\footnote{\url{https://github.com/alexandergagliano/Prost}} \citep{alex_gagliano_2025_15635013}, which calculates the posterior probability that each galaxy in a given search region is the true host galaxy using the fractional offset, redshift, and brightness of the host/transient. For this analysis, we used the galaxy catalog from the Pan-STARRS DR2 \citep{Chambers2017yCat.2349....0C}. We notice that there are several extended faint sources are visible near the SN (see the right panel of Figure \ref{fig:sn_image}). These sources are very faint and are not listed in the Pan-STARRS catalog.
To evaluate whether they are associated with the host galaxy CGCG~505-052 (and would, therefore, be probable birth sites of the SN), we use $gri$-band images from Gemini-North around 76 days after peak, while the SN was still visible in the redder bands, and the seven extended sources are marked in the right panel of Figure \ref{fig:sn_image}. The $gri$-band photometry of these sources are fitted with Prospector \citep{Leja2019ApJ...877..140L,Johnson2021ApJS..254...22J}, using the same model and method as discussed in \citet{nugent2022}, and the inferred redshifts are indicated in Figure~\ref{fig:sn_image}.
Although the inferred redshifts are generally much higher than $z = 0.027$, suggesting they may be background galaxies, the large uncertainties inherent in photometric redshift estimates prevent us from definitively ruling out the possibility that some of them are satellite galaxies at the same redshift as CGCG~505-052, and thus potential birth sites of SN~2024acyl.

}

\corr{To understand where SN~2024acyl stands in the Type Ibn population, we compare the projected offset and the host-normalized offset (i.e., normalized by the host galaxy's half-light radius) of SN2024acyl with a group of SNe Ibn. SN~2024acyl is $\sim$59.6$\arcsec$ away from the center of CGCG~505-052, which has a half-light radius of 12.8\arcsec, corresponding to a projected physical distance of $\sim$35.3~kpc. 
We obtain the 66 SNe that are classified as Type Ibn on TNS up to 2025 April 1. The host associations for these objects are determined using \texttt{Pr\"ost}.}
We plot the projected offset distribution of Type Ibn SNe in the upper panel of Figure \ref{fig:host_offset}, where the black vertical line marks SN~2024acyl. SN~2024acyl has the largest projected host offset among all Type Ibn SNe.
In the bottom panel of Figure \ref{fig:host_offset}, we show the cumulative distribution of host offsets normalized by their host half-light radius. The host half-light radius are taken from the DECam Legacy Survey \citep[DECaLS;][]{Dey2019AJ....157..168D} catalog, or from the the Sloan Digital Sky Survey \citep[SDSS;][]{York2000AJ....120.1579Y} catalog if the object is not in the DeCaLS catalog. SN~2024acyl has one of the largest normalized offsets ($\sim$4.7) among all Type Ibn SNe, \corr{which also places it in the top 0.5\% of the distribution for core-collapse SNe.}

To determine the global properties of the host galaxy, we jointly model the photometry and spectroscopy of SN~2024acyl's host, CGCG~505-052, using the stellar population inference code Prospector \citep{Leja2019ApJ...877..140L,Johnson2021ApJS..254...22J}, following the methodology described in \cite{Nugent2025ApJ...982..144N}, \corr{which employs a seven-bin non-parametric star formation history model.} The host global aperture photometry \corr{($ugrizJHK$)} is taken from Blast \corr{\citep{Jones2024arXiv241017322J}}\footnote{https://blast.scimma.org/}, where the global aperture is determined from the $g$-band image using Photutils \citep{larry_bradley_2022_6825092}.
The Prospector fit yields a stellar mass of $\rm log(M_{*})/M_{\odot} = 10.60^{+0.04}_{-0.04}$, a SFR of $\rm 1.52^{+0.47}_{-0.30}\,M_{\odot}\,yr^{-1}$, and a stellar metallicity of $\rm log(Z_{*}/Z_{\odot}) = 0.03^{+0.05}_{-0.04}$. The stellar mass and metallicity of the SN~2024acyl host are consistent with those of other SNe Ibn hosts reported in
\cite{Qin2024MNRAS.533.3517Q}, while its SFR lies on the higher end of their sample. We emphasize that this comparison refers only to the global host properties, and the local environment of SN~2024acyl will be discussed in the next section.



\begin{figure}
\includegraphics[width=1.\linewidth]{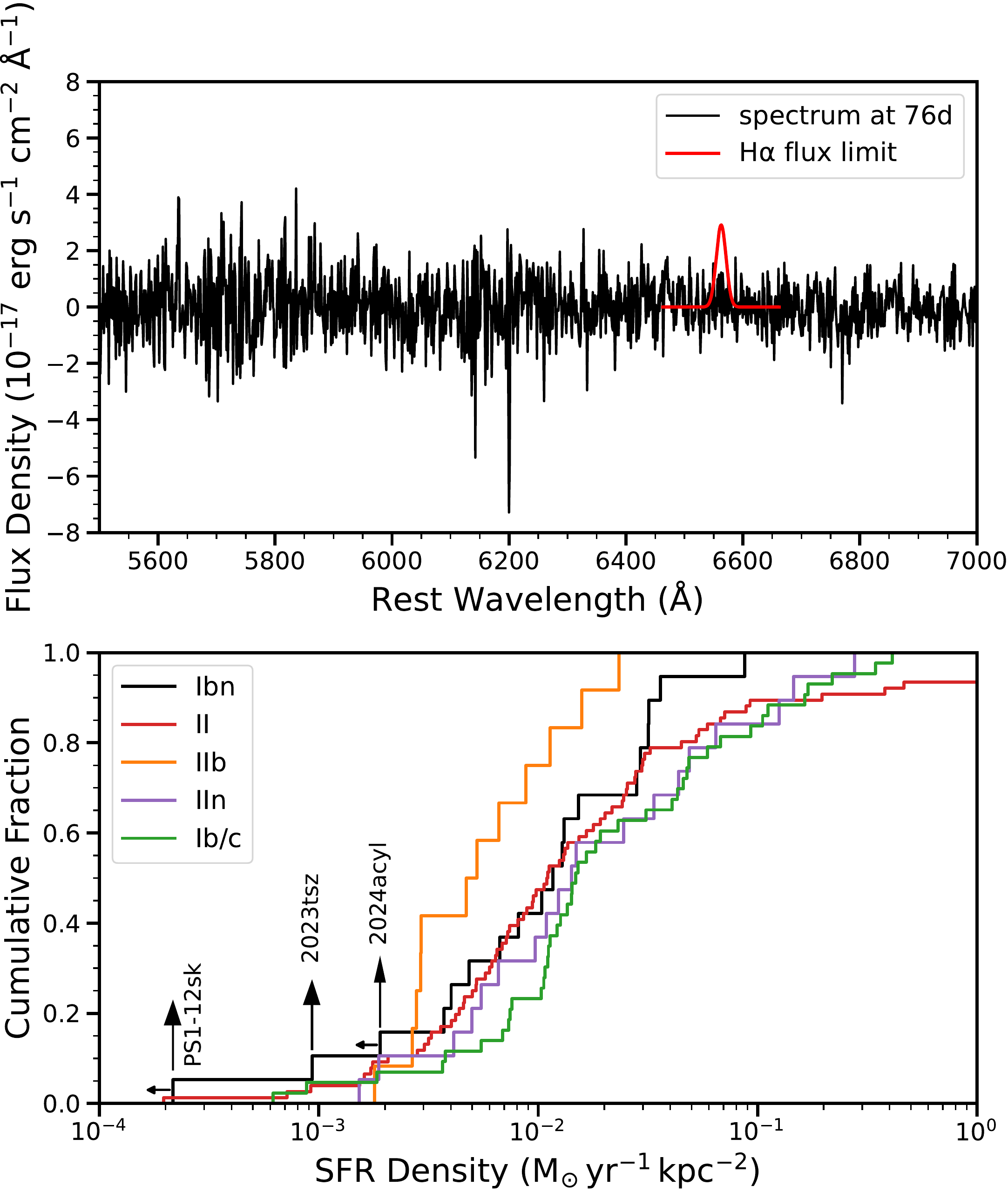}
\caption{Upper: Spectrum taken 76 days after maximum light at the position of SN~2024acyl. We constrain the upper limit on the H$\alpha$ luminosity by implanting a simulated emission line (red) with a flux equal to four times the rms of the continuum and a FWHM of 1000~$\rm km\,s^{-1}$. The spectrum has been flux calibrated and corrected for extinction. Bottom: The SFR surface density of SNe Ibn \citep{Hosseinzadeh2019ApJ...871L...9H} and other core-collapse SNe types \citep{Galbany2018ApJ...855..107G}. The SFR density upper limit for SN~2024acyl and PS1-12sk, and the SFR density of SN~2023tsz are indicated by black arrows.
\label{fig:ha_limit}}
\end{figure}

\subsection{Local Host Environment}
\label{sec:local_host}
To investigate whether there is any faint \corr{underlying star cluster} behind SN~2024acyl, we stacked archival $w$- and $i$-band images from PSST using Swarp \citep{Bertin2002}. The zero points of the stacked $w$- and $i$-images were calibrated to $r$ and $i$ bands, respectively, using the PS1 catalog. The PSF photometry were performed at the position of SN~2024acyl using Photutils \citep{larry_bradley_2022_6825092}. There is no source detected down to 24.5~mag in the $w$ band and 22.3~mag in the $i$ band, which translates to an absolute magnitude of $\rm M_{r}>-11.2$~mag and $\rm M_{r}>-13.3$~mag \corr{assuming the distance of SN~2024acyl}.

To further constrain the star formation rate at the explosion site of SN~2024acyl, we analyze a deep \corr{long-slit} spectrum (6,000 s exposure) taken with Gemini-North 76 days after peak brightness. We refer the reader to Appendix~\ref{append:data_reduc} for details on data acquisition and reduction of this spectrum. As shown in Figure~\ref{fig:ha_limit}, no clear SN signal is detected in the spectrum. To place a constraint on any undetected H$\alpha$ emission, we follow the methodology described in \cite{Sand2021ApJ...922...21S}, implanting a simulated H$\alpha$ emission line into the spectrum (upper panel of Figure~\ref{fig:ha_limit}). The spectral resolution in this wavelength range, measured from the full width at half maximum (FWHM) of sky lines, is approximately 820~$\rm km\,s^{-1}$. For the implanted H$\alpha$ line, we conservatively assume a FWHM of 900~$\rm km\,s^{-1}$. The peak flux is set to four times the rms of the observed spectrum, following \cite{Sand2021ApJ...922...21S}. This results in an H$\alpha$ flux limit of $\rm 6.1\times10^{-16}\ erg\,s^{-1}\,cm^{-2}$, which corresponds to a luminosity limit of $\rm 1.1\times10^{38}\ erg\,s^{-1}$ at a distance of 122~Mpc.
We estimate the SFR using the conversion factor from
\cite{Kennicutt1998ARA&A..36..189K}: $\rm SFR (M_{\odot}\,yr^{-1}) = 7.9 \times 10^{-42} L_{H\alpha}(erg\,s^{-1})$, which yields an upper limit on the SFR of $\rm 8.6\times10^{-4}\ M_{\odot}\,yr^{-1}$.

To enable a direct comparison with the SFR surface density estimated in \cite{Hosseinzadeh2019ApJ...871L...9H} for a sample of Type Ibn SNe, we estimate the local SFR surface density at the site of SN~2024acyl using \corr{the} SFR we derived above from the long slit data. The slit width used to obtain the spectrum was 1.5\arcsec, which corresponds to a projected distance of $\sim$0.9~kpc at the distance of the SN. The spatial FWHM we used to extract the trace on the 2D spectrum was around 0.9\arcsec, corresponding to a projected distance of $\sim$0.5~kpc. We then estimate the upper limit on the SFR surface density at the location of SN~2024acyl: $\rm \Sigma_{SFR} = 8.6\times10^{-4}\ M_{\odot}\,yr^{-1}/(0.9~kpc \times 0.5~kpc) = 1.9\times10^{-3}\ M_{\odot}\,yr^{-1}\,kpc^{-2}$. \corr{This estimate assumes that the potential host source is at least as extended as the slit width, and we emphasize that this value should be considered as an order-of-magnitude approximation.}
In the bottom panel of Figure \ref{fig:ha_limit}, we show the SFR densities at the locations of a sample of SNe Ibn \citep{Hosseinzadeh2019ApJ...871L...9H} and Type Ibn SN~2023tsz \citep{Warwick2025MNRAS.536.3588W}, along with other types of core-collapse SNe \citep{Galbany2018ApJ...855..107G}. The upper limit we derived makes SN~2024acyl the third Type Ibn SN (after PS1-12sk and SN~2023tsz) discovered in a low SFR region.

\subsection{Implications from Host Environments}
Type Ibn SNe are generally thought to be the explosions of stripped massive stars in He-rich environments \citep[e.g.,][]{Pastorello2007Natur.447..829P,Pastorello2015MNRAS.449.1954P}, and thus are expected to be associated with star-forming regions. However, this has been challenged by the discovery of PS1-12sk \citep{Sanders2013ApJ...769...39S}, in which the object was found in a region with extremely low SFR \citep{Hosseinzadeh2019ApJ...871L...9H}. SN~2023tsz was also found in a low-mass galaxy with a low SFR  \citep{Warwick2025MNRAS.536.3588W}. Its host has an absolute magnitude of $M_{r} = -13.2$. A source of similar brightness, if present at the site of SN~2024acyl, would have been detected.

\corr{SN~2024acyl is located at a projected distance of around 10~kpc from the the nearest significant light of its host galaxy (see Figure \ref{fig:sn_image})}. Assuming a massive star progenitor with a lifetime of 10~Myr (typical for a 15 $\rm M_{\odot}$ star, \citealt{Schaller1992A&AS...96..269S}), the velocity required to reach the explosion site is roughly \corr{given by the offset distance divided by the travel time:} \corr{$\rm D_{offset}/t_{travel}$ = $\rm 980 \times (D/10kpc)/(t_{*}/10Myr)\ km\,s^{-1}$}. As discussed in Section \ref{sec:global_host}, we cannot fully exclude the possibility that one of the \corr{seven} extended sources near SN~2024acyl is its birth site. \corr{The nearest of these sources, in terms of angular separation, is at a projected distance of 3.5~kpc if we assume it is associated with the SN, which would imply a progenitor velocity of $\sim$340~$\rm km\,s^{-1}$.} 
It is possible that the progenitor of SN~2024acyl was a runaway star, ejected from its birth site either through close encounters in a dense cluster or by the SN explosion of a primary star in a massive binary system \citep{Eldridge2011MNRAS.414.3501E}.
However, such velocities are unusually high for runaway stars massive enough to explode as core-collapse SNe \citep[e.g.,][]{Hoogerwerf2001A&A...365...49H,Eldridge2011MNRAS.414.3501E,Renzo2019A&A...624A..66R}.

Another possibility is that the progenitor of SN~2024acyl is not a very massive star, as has been discussed for PS1-12sk \citep{Hosseinzadeh2019ApJ...871L...9H}. For example, a He detonation on the surface of a white dwarf within a He-rich CSM could produce Type Ibn SNe. Such a scenario would require a long delay time and may explain the large host offsets and low SFR densities observed in both SN~2024acyl and PS1-12sk. 
\corr{Based on optical and UV observations of SNe Ibn, \cite{Dessart2022A&A...658A.130D} and \cite{Wang2024MNRAS.530.3906W} suggested that the progenitor systems of some Type Ibn SNe are likely low-mass He stars in binary systems. In this case, the lifetime of the He star progenitor could be longer. Moreover, if the companion star has already undergone a SN explosion, the surviving binary may have received a significant kick velocity, further contributing to the large offset from its birth site.
}\cite{Ko2025MNRAS.tmp.1187K} proposed a progenitor channel for Type Ibn SNe in which both stars in the binary system are initially below 8~$\rm M_{\odot}$. During binary interaction, the initially less massive star accretes material from the primary, becomes above 8~$\rm M_{\odot}$, and is stripped to a He star in the final-stage evolution. This secondary then eventually explodes as a core-collapse SN. This pathway may involve long delay times and produce large offsets from star-forming regions.
Alternatively, there could be faint, undetected star-forming regions at the explosion site, consistent with a massive star progenitor, which would require telescopes such as \corr{the \textit{Hubble Space Telescope}} to confirm.

The large host offset and low local SFR of SN~2024acyl add to the mystery surrounding the progenitor systems of Type Ibn SNe. A systematic study at the explosion sites of Type Ibn SNe would offer valuable insights into the nature of their progenitors.



\begin{figure*}
\includegraphics[width=1.\linewidth]{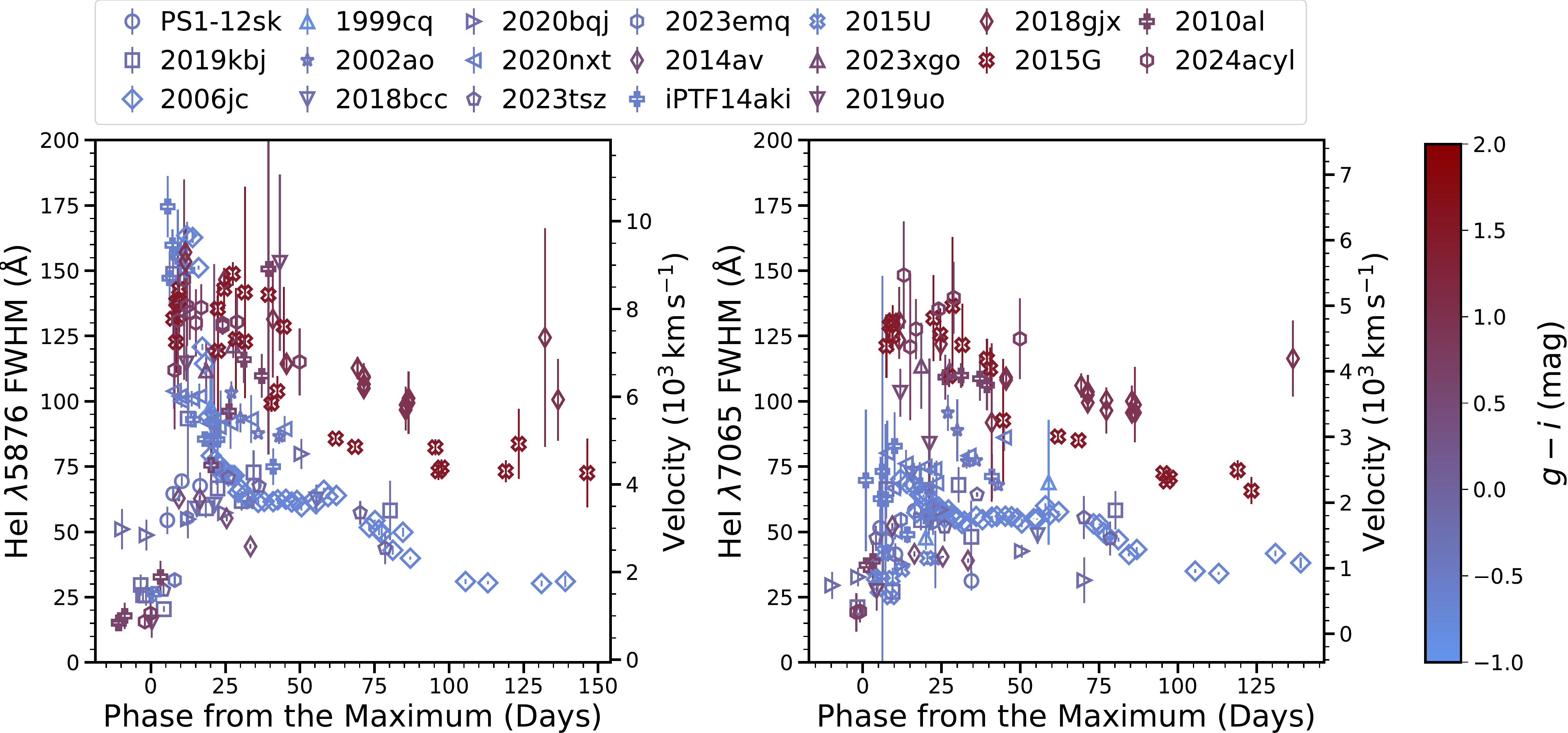}
\caption{The FWHM evolution of the He~I $\lambda5876$ (Left) and He~I $\lambda7065$ (Right) of SNe Ibn. The color represents the average $g-i$ color index $\sim20-35$ days after peak.
\label{fig:fwhm_evol}}
\end{figure*}

\section{Spectral Diversity in Type Ibn SNe} \label{sec:spec_diversity}

While the light curves of SNe Ibn are largely homogeneous, typically exhibiting a fast rise followed by a rapid decline \citep{Pastorello2016MNRAS.456..853P,Hosseinzadeh2017ApJ...836..158H}, their spectral evolution near peak brightness displays notable diversity (\citealt{Hosseinzadeh2017ApJ...836..158H}, hereafter H17). Around maximum light, some SNe Ibn exhibit narrow P-Cygni profiles, whereas others are dominated by broader emission features. 
This diversity could result from differences in CSM optical depth, viewing angles, or even distinct progenitor channels (H17).

\subsection{Spectral Diversity in SNe Ibn \corr{A Month} After Maximum}
As discussed in Section \ref{sec:spec_evol}, around 20--35 days after peak, when the spectra are no longer dominated by hot blackbody continua, the SNe Ibn in our comparison sample separate into two groups: one that shows strong blue \corr{continuum} emission and narrower He~I lines, and another that lacks prominent blue emission and exhibits broader He~I lines. 

\corr{To investigate this further, we select Type Ibn SNe with spectral coverage between 20 and 35 days after peak, yielding 19 objects in total. Redshifts as well as Galactic and host extinction values are adopted from the references listed in Table~\ref{tab:snibn_properties}. 
}
We measure the FWHM of the two most prominent He lines, He~I $\lambda5876$ and He~I $\lambda7065$ lines, by fitting a \corr{single} Gaussian profile to the emission lines, with the continuum defined by a line connecting two local minima. 
We also calculated the synthetic $g-i$ color from redshift-corrected spectra in this phase range, ensuring that the comparison is not biased by redshift effects. \corr{We focus on the $g$ and $i$ bands, as $g$ spans the wavelengths of the blue continuum, and $i$ is not affected by the blue continuum at these phases (see Figure~\ref{fig:spec_comp}).}

Figure \ref{fig:fwhm_evol} shows the temporal evolution of the He~I $\lambda5876$ and He~I $\lambda7065$ line widths, color-coded by the average $g-i$ color between 20 and 35 days after peak. 
The FWHM values increase in the early phases and slowly decline afterwards. The corresponding velocities are generally consistent with those reported in previous studies \citep{Pastorello2016MNRAS.456..853P,Hosseinzadeh2017ApJ...836..158H,Gangopadhyay2020ApJ...889..170G}.
Notably, we find that objects with \corr{narrower} lines tend to show bluer \corr{synthetic $g-i$ color}, a trend that is especially clear in the He~I $\lambda7065$ line.

\begin{figure}
\includegraphics[width=1.\linewidth]{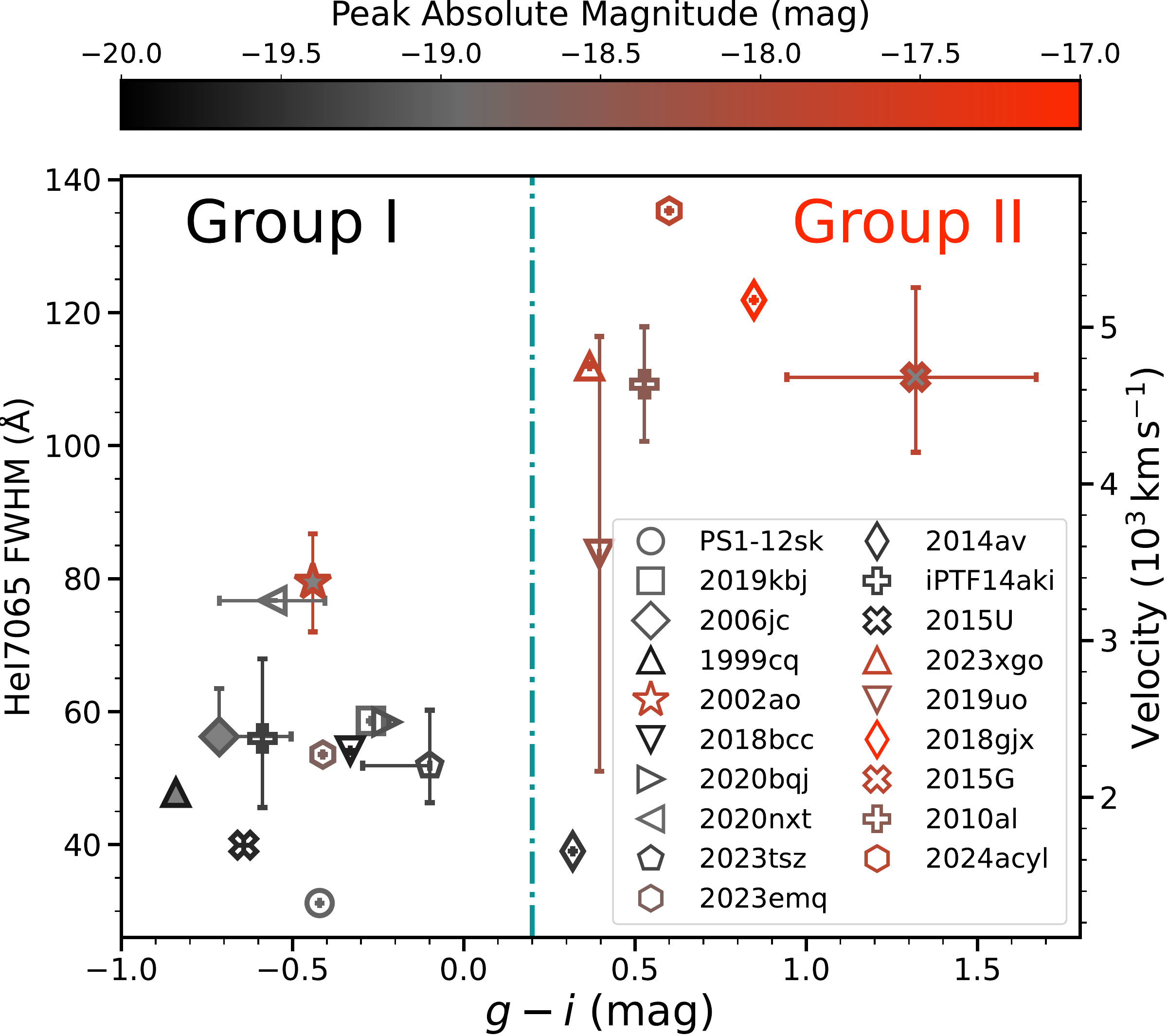}
\caption{The average $g-i$ color versus FWHM of the HeI $\lambda7065$ line between 20 and 35 days after maximum for a group of SNe Ibn. The peak magnitude of each object is indicated by color, with darker tones representing brighter objects and redder tones representing fainter ones. All magnitudes are in the AB system. 
The objects appear to separate into two groups: Group I ($g-i \lesssim 0.2$) and Group II ($g-i \gtrsim 0.2$). Group I events generally exhibit brighter peak magnitudes and narrower He~I lines.
\label{fig:color_fwhm_comp}}
\end{figure}

In Figure \ref{fig:color_fwhm_comp}, we compare the average FWHM \corr{(weighted by the uncertainties of FWHM)} of the He~I $\lambda7065$ line to the $g-i$ color index measured between 20 and 35 days for these objects. We focus on He~I $\lambda7065$ because it is both prominent and relatively isolated, whereas He~I $\lambda5876$ is likely blended with the Na {\sc i} doublet \citep[e.g.,][]{Pastorello2016MNRAS.456..853P}. To estimate systematic uncertainties introduced by uncertainty in the peak epoch ($t_{\rm peak}$), we shifted the measurement window forward and backward by the estimated error in $t_{\rm peak}$. The variation in the measured values is adopted as the systematic uncertainty. 
In addition, we explore whether the trend we identified is related to the light curve peak brightness.   
We indicate the peak brightness of each object using color, with darker tones representing brighter objects and redder tones representing fainter ones. 
Surprisingly, from Figure \ref{fig:color_fwhm_comp}, we see a clear trend that bluer objects tend to have narrow He lines as well as higher peak brightness. 
We list the sample we analyzed as well as the parameters we measured in Table \ref{tab:snibn_properties}. 

\begin{figure}
\includegraphics[width=1.\linewidth]{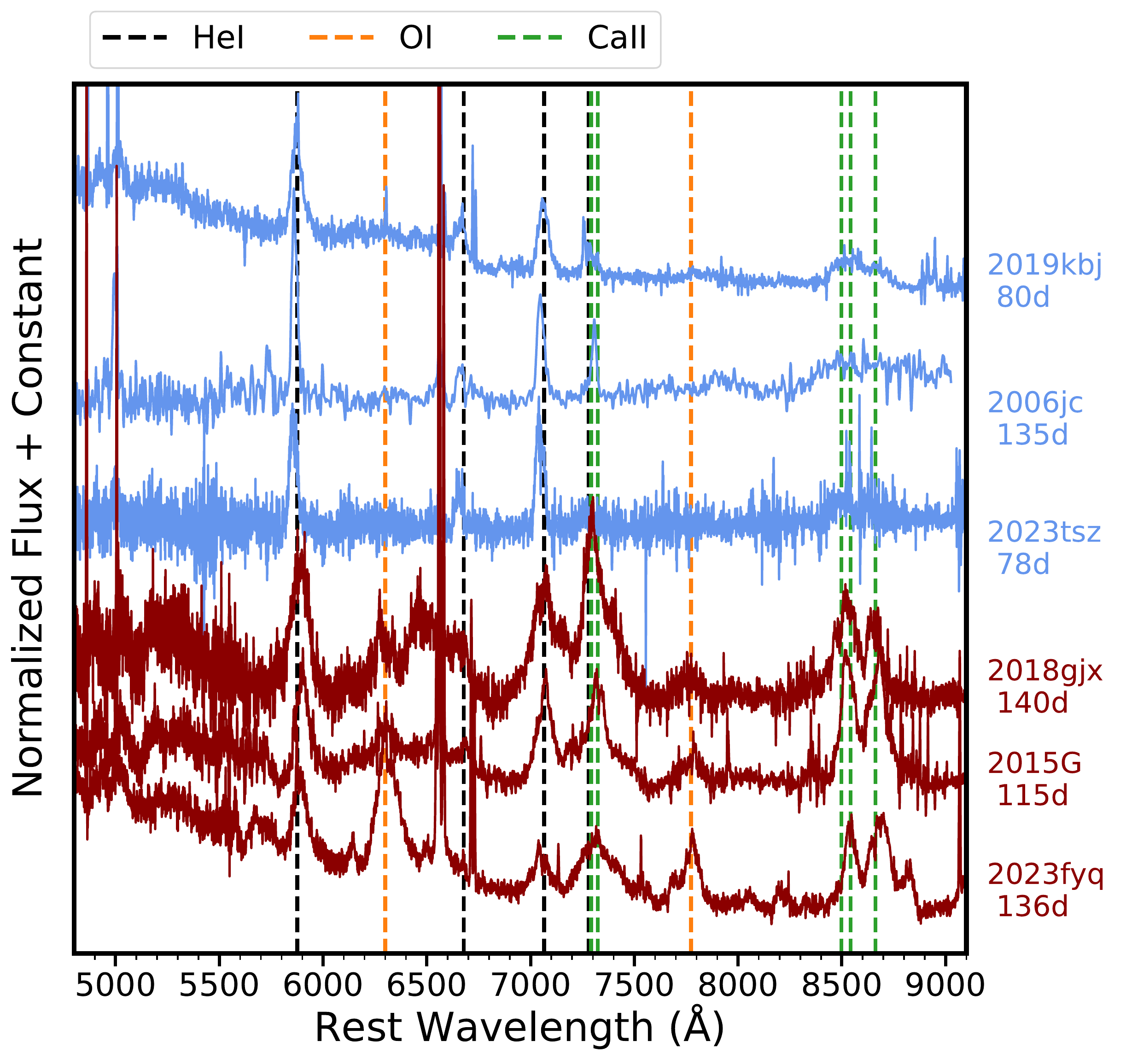}
\caption{Comparison of late-time ($\sim$80--130 days after peak) spectra for Type Ibn SNe with available observations at these phases. \corr{Group~I objects (SN~2019kbj, SN~2006jc, and SN~2023tsz) are marked in blue, while Group~II objects (SN~2018gjx and SN~2015G) are shown in red. The color coding of these five objects is consistent with that used in Figure \ref{fig:spec_comp}. The late-time spectrum of SN~2023fyq closely resembles those of SN~2018gjx and SN~2015G and is therefore also marked in red.}
\label{fig:nebular_spec_comp}}
\end{figure}

\corr{In Figure \ref{fig:color_fwhm_comp}, the objects in the sample appear to separate into two groups. For convenience, we define Group~I as objects with $g - i < 0.2$ and  Group~II as those with $g - i > 0.2$. This division is arbitrarily chosen based on Figure \ref{fig:color_fwhm_comp}.} Group I objects tend to show narrower He lines ($\rm FWHM_{HeI7065} < \sim80~\text{\AA}$ or $v_{\rm HeI7065} < 3300~\rm km\,s^{-1}$) and higher peak luminosities, whereas Group II objects tend to show broader He lines ($\rm FWHM_{HeI7065} > 80~\text{\AA}$ or $v_{\rm HeI7065} > 3300~\rm km\,s^{-1}$) and lower peak luminosities. \corr{The bluer color is associated with blue continuum emission arising from the Fe~II forests (see Figure \ref{fig:spec_comp} for an example). Notably, SN~2002ao appears to be an exception to this trend, and SN~2014av is a mild outlier.} We further discuss the classification of these two objects in Section \ref{sec:connections_to_h17}.

Interestingly, this division between two groups likely persist into later phases. In Figure \ref{fig:nebular_spec_comp}, we compare the spectra taken at around 80 to 130 days after the peak light for those objects in Figure \ref{fig:spec_comp} that have late-time observations, where objects marked with blue and red belong to Group I and Group II, respectively.
At this late phase, Group I continues to display predominantly narrow He~I lines, whereas Group II shows broader He lines and relatively prominent oxygen emission features. In particular, the [OI]$\lambda\lambda6300,\, 6364$ doublet is much weaker in Group I, while Group II displays a range of line strengths, which has been attributed to a viewing-angle effect, possibly due to the asymmetric distribution of CSM around the progenitor \citep{Dong2024ApJ...977..254D}. Additionally, the Ca~II lines in Group II appear broader and more prominent than in Group I. 

It is worth noting that the three Type Ibn SNe with atypical host environments (SN~2024acyl, PS1-12sk, and SN~2023tsz) do not fall into the same group: PS1-12sk and SN~2023tsz are clearly in Group I, whereas SN~2024acyl belongs to Group II. A systematic study of Type Ibn host galaxies will be needed to fully examine the diversity of their environments. \corr{In addition, the connection between pre-explosion activity and the spectral diversity we identify remains unclear. Both SN~2006jc (Group~I) and SN~2019uo (Group~II) show short-duration, outburst-like precursors one to two years before explosion. While SN2023fyq lacks spectral coverage between 20 and 35 days, its similarity to SN2015G at both early \citep{Dong2024ApJ...977..254D} and very late phases (see Figure \ref{fig:nebular_spec_comp}) suggests it most likely belongs to Group~II. Notably, SN2023fyq showed a long-duration precursor event \citep{Brennan2024A&A...684L..18B,Dong2024ApJ...977..254D}, which differs significantly from the short-lived outbursts seen in SN~2006jc and SN~2019uo.
}




\begin{deluxetable}{ccccccc} \label{tab:snibn_properties}
\tabletypesize{\scriptsize}
\tablecaption{Observation Properties of SNe Ibn in our sample}
\tablewidth{0pt}
\tablehead{
\colhead{Name} &
\colhead{$g-i$$^a$} &
\colhead{$\rm FWHM_{7065}$$^b$} &
\colhead{$M_{\rm peak}$} & 
\colhead{Subtype} &
\colhead{Subtype (H17)$^c$} &
\colhead{Reference} 
}
\startdata 
1999cq$\ddag$ & $-0.8$ & $47.8_{-4.7}^{+4.7}$ & $-19.93 (R)$ & Group I? & P-Cygni? & \cite{Matheson2000AJ....119.2303M}\\
2002ao$\ddag$ & $-0.4$ & $79.5_{-7.5}^{+7.3}$ & $-18.04 (R)$ & ? & P-Cygni? & \cite{Foley2007ApJ...657L.105F}, \cite{Pastorello2008MNRAS.389..113P}\\
2006jc$\ddag$ & $-0.7^{+0.2}_{-0.1}$ & $56.2_{-1.1}^{+7.2}$ & $-19.32 (R)$ & Group I & Emission & \cite{Foley2007ApJ...657L.105F}, \cite{Pastorello2008MNRAS.389..113P}\\
PS1-12sk$\ddag$ & $-0.4$ & $31.2_{-3.7}^{+3.7}$ & $-19.00 (z)$ & Group I & Emission & \cite{Sanders2013ApJ...769...39S}\\
iPTF14aki$\ddag$ & $-0.6$ & $56.0_{-10.4}^{+12.0}$ & $-19.56 (R)$ & Group I & Emission & H17\\
2015U$\ddag$ & $-0.6$ & $39.9_{-2.6}^{+2.6}$ & $-19.61 (r)$ & Group I & Emission & H17\\
2014av$\ddag$ & $0.3$ & $39.0_{-0.8}^{+0.8}$ & $-19.64 (R)$ & ? & ? & \cite{Pastorello2016MNRAS.456..853P}\\
2018bcc & $-0.3$ & $54.1_{-3.5}^{+3.5}$ & $-19.68 (r)$ & Group I & Emission & \cite{Karamehmetoglu2021AA...649A.163K}\\
2019kbj & $-0.3^{+0.1}_{-0.1}$ & $58.6_{-3.1}^{+3.1}$ & $-18.99 (r)$ & Group I & Emission & \cite{Ben-Ami2023ApJ...946...30B}, \cite{Dong2024ApJ...977..254D}\\
2020bqj & $-0.2$ & $58.4_{-2.7}^{+2.7}$ & $-19.23 (r)$ & Group I & Emission & \cite{Kool2021AA...652A.136K}\\
2020nxt & $-0.6^{+0.2}_{-0.2}$ & $76.7_{-1.9}^{+1.8}$ & $-18.95 (o)$ & Group I & Emission & \cite{Wang2024MNRAS.530.3906W}\\
2023emq & $-0.4$ & $53.5_{-2.0}^{+2.0}$ & $-18.70 (o)$ & Group I & Emission & \cite{Pursiainen2023ApJ...959L..10P}\\
2023tsz & $-0.1^{+0.0}_{-0.2}$ & $51.9_{-5.6}^{+8.3}$ & $-19.70 (g)$ & Group I & Emission & \cite{Warwick2025MNRAS.536.3588W}\\
2010al$\ddag$ & $0.5$ & $109.2_{-8.6}^{+8.6}$ & $-18.56 (R)$ & Group II & P-Cygni & \cite{Pastorello2015MNRAS.449.1921P,Pastorello2016MNRAS.456..853P}\\
2015G$\ddag$ & $1.3^{+0.4}_{-0.4}$ & $110.3_{-11.3}^{+13.5}$ & $-17.89 (r)$ & Group II & P-Cygni? & \cite{Shivvers2017MNRAS.471.4381S}, H17\\
2018gjx & $0.8$ & $121.9_{-4.3}^{+4.3}$ & $-17.10 (r)$ & Group II & P-Cygni & \cite{Prentice2020MNRAS.499.1450P}\\
2019uo & $0.4$ & $83.7_{-32.7}^{+32.7}$ & $-18.30 (r)$ & Group II & P-Cygni & \cite{Gangopadhyay2020ApJ...889..170G}\\
2023xgo & $0.4$ & $111.9_{-4.3}^{+4.3}$ & $-17.65 (r)$ & Group II & P-Cygni & this work; \cite{Gangopadhyay2025MNRAS.tmp.1456G}\\
2024acyl & $0.6$ & $135.3_{-3.1}^{+3.1}$ & $-17.90 (r)$ & Group II & P-Cygni & this work\\
\enddata
\tablenotetext{$\ddag$}{Objects also analyzed in H17.}
\tablenotetext{$a$}{The $g-i$ color is measured from synthetic photometry of the spectra. We do not estimate statistical uncertainties for $g-i$; the quoted errors reflect systematic uncertainties.}
\tablenotetext{$b$}{The weighted average FWHM of the $\rm He~I\,\lambda7065$ line measured between 20 and 35 days after peak brightness.}
\tablenotetext{$c$}{Subtype classification based on H17. For objects already classified in H17, we adopt their labels; otherwise, the classification is based on data from the listed references.}
\end{deluxetable}

\begin{deluxetable}{lcc}
\tablecaption{Summary of the main characteristics of Group I vs.\ Group II\label{tab:two_group}}
\tablewidth{0pt}
\tablehead{
\colhead{Phase} & \colhead{Group I} & \colhead{Group II}
}
\startdata
$\sim$0 days & He emission lines & He P-Cygni profiles \\
20--35 days  & Bluer color; narrower He\,I lines; higher peak brightness & Redder color; broader He\,I lines; lower peak brightness \\
$\gtrsim$80 days & Narrow He lines; weak or absent [O\,I] & Broad He lines; [O\,I] with varying strength \\
\hline
Additional diagnostics & Never show inner ejecta & Show inner ejecta around one month after peak \\
\hline
\multicolumn{3}{c}{\hspace{6em}\textbf{Possible interpretations}} \\
 & lower explosion energy    & higher explosion energy   \\
 & and/or a higher-density, more-extended CSM & and/or a lower-density, less-extended CSM \\
\enddata
\end{deluxetable}

\subsection{Connections To The Spectral Diversity Presented In \cite{Hosseinzadeh2017ApJ...836..158H}} \label{sec:connections_to_h17}
H17 proposed two spectral subclasses of SNe Ibn based on He~I line profiles around maximum light: one group showing narrow P-Cygni profiles, and another with broader emission lines. The diversity they identified is closely connected to the diversity we report here. In Table \ref{tab:snibn_properties}, We indicate the group assignment for each object based on this work as well as the classification scheme in H17.
Most P-Cygni subclass objects fall into our Group II, while most emission-dominated objects fall into Group I, with a few exceptions. 
For SN~1999cq and SN~2002ao, no observations were obtained around peak brightness, making it unclear whether they exhibited P-Cygni features. Their exact peak epochs are also uncertain, which makes it difficult to classify them within either framework. SN~2014av shows narrow P-Cygni absorption superimposed on broader emission lines near peak, so H17 excluded it from both subclasses. In our analysis, SN~2014av is also difficult to classify, as it exhibits both a large $g-i$ color and a narrow $\rm He~I\,\lambda7065$ line, placing it ambiguously between the two groups.

The diversity from peak light to late phases in Type Ibn SNe convey information about the CSM and the progenitor properties. For example, the FWHM of the He~I lines roughly reflects the expansion velocity of the line-forming region, while the peak brightness is likely influenced by the degree of CSM interaction.
We summarize the main characteristics of two groups in Table \ref{tab:two_group}.
In the next subsection, we will discuss how the diversity we identify can be related to the CSM and progenitor properties of Type Ibn SNe.

\subsection{Implications For The Progenitors of SNe Ibn}

We hypothesize that the diversity observed between the two groups likely reflects differences in CSM configurations. In Group I, narrow He emission lines dominate the spectra from early to late times. At early phases, \corr{He lines undergo electron scattering} in the CSM above the photosphere, producing narrow emission with relatively broad wings. At later times, those broad wings become less prominent as the optical depth of the CSM decreases, while the narrow emission lines persist. This behavior resembles that of some Type IIn SNe, such as SN~1999W, where narrow H$\alpha$ emission persists throughout the evolution \citep{Sollerman1998ApJ...493..933S,Chugai2004MNRAS.352.1213C,Dessart2016MNRAS.458.2094D}. The persistent absence of broad lines suggests that the ejecta is effectively decelerated by strong CSM interaction, with most of the material piling up in a slow-moving cold dense shell \citep{Dessart2016MNRAS.458.2094D}. This scenario may result from a weak explosion and/or a particularly dense \corr{and extended} CSM \citep{Dessart2016MNRAS.458.2094D,Dessart2022A&A...658A.130D}. The presence of strong Fe~II line forests, \corr{most likely originating from the cold dense shell formed during CSM interaction \citep{Chugai2009MNRAS.400..866C, Dessart2022A&A...658A.130D}}, supports this interpretation and indicates relatively strong interactions. Moreover, the relatively high peak brightness observed in Group I may reflect more efficient conversion of kinetic energy into radiation due to the strong deceleration of the ejecta.

In contrast, Group II displays narrow P-Cygni lines from unshocked CSM at early times, followed by the emergence of broad lines. This transition, reminiscent of some Type IIn/IIP SNe such as SN~1998S \citep{Fassia2001MNRAS.325..907F, Leonard2000ApJ...536..239L, Dessart2016MNRAS.458.2094D} and SN~2020pni \citep{Terreran2022ApJ...926...20T}, indicates that the ejecta have swept up most of the CSM, allowing the photosphere to recede into the inner ejecta. These features point to a less dense and \corr{less extended CSM} and/or a more energetic explosion in Group II. 
This picture is consistent with the relatively weak Fe~II lines weeks after peak and lower peak brightness observed in Group II, both of which point to less efficient conversion of kinetic energy into radiation.
At very late times, Group II objects enter the nebular phase, and their spectra resemble those of normal Type Ib/c SNe \citep{Dong2024ApJ...977..254D}, apart from the persistent He emission lines. This indicates that both ongoing CSM interaction and emission from the inner ejecta are present simultaneously, possibly due to an asymmetric CSM geometry shaped by pre-explosion activity \citep{Dong2024ApJ...977..254D}. \corr{However, it is worth noting that given the current observations we have, it is not clear how the pre-explosion activity is connected to the diversity we identify. 
}

\corr{Potential} progenitors of SNe Ibn have largely fallen into two categories: 1) a massive Wolf-Rayet star explodes within He-rich CSM, produced by pre-SN stellar winds or eruptions of the massive star \citep{Pastorello2007Natur.447..829P,Foley2007ApJ...657L.105F,Pastorello2016MNRAS.456..853P,Woosley2017ApJ...836..244W,Hosseinzadeh2017ApJ...836..158H}; and 2) a binary system that contains a He star, in which the surrounding CSM originates from binary interactions \citep{Pastorello2007Natur.447..829P,Foley2007ApJ...657L.105F,Maund2016ApJ...833..128M,Pastorello2016MNRAS.456..853P,Sun2020MNRAS.497.5118S,Wu22,Metzger22,Sun2023MNRAS.521.2860S,Wang2024MNRAS.530.3906W,Tsuna2024OJAp....7E..82T,Dong2024ApJ...977..254D,Pellegrino2024ApJ...977....2P,Ercolino2025A&A...696A.103E,Ko2025MNRAS.tmp.1187K,Ercolino2025arXiv251004872E}. In the case of PS1-12sk, due to the low SFR density at its explosion site, the probability that some SNe Ibn originate from white dwarf explosions (He detonations) were also discussed in \cite{Sanders2013ApJ...769...39S} and \cite{Hosseinzadeh2019ApJ...871L...9H}. 

The diversity in the progenitor systems of SNe Ibn may lead to a range of CSM properties, contributing to the observed diversity discussed in this paper. Pre-explosion mass ejections are expected from some massive stars during the late stages of nuclear burning in their cores \citep{Quataert2012, Shiode2014, Fuller2017, Fuller2018MNRAS.476.1853F, Morozova2020}. For He stars with masses in the range of 2.5-3.2~$\rm M_{\odot}$, eruptive mass loss is anticipated, potentially triggered by final-stage silicon deflagration or detonation in the core prior to the SN explosion \citep{Woosley2019ApJ...878...49W,Ertl2020ApJ...890...51E,Moriya2025arXiv250705506M}. Depending on the timing and energy of the silicon flash, a range of ejected mass is possible \citep{Woosley2019ApJ...878...49W,Ertl2020ApJ...890...51E,Moriya2025arXiv250705506M}, which may in turn lead to diverse CSM properties and corresponding observational signatures.

For the binary progenitor scenario, late-time interactions between a He star and either a main sequence star or a compact object can induce high-rate mass transfer, leading to the ejection of a substantial amount of He-rich CSM prior to explosion \citep{Wu22,Takei2024ApJ...961...67T,Ercolino2025A&A...696A.103E,Tsuna2024OJAp....7E..82T,Dong2024ApJ...977..254D,Ercolino2025A&A...696A.103E,Ercolino2025arXiv251004872E}. Variations in He star mass, orbital period, and the nature of the companion (compact object vs. main sequence star) can produce a wide range of CSM masses and spatial extents \citep{Wu22,Takei2024ApJ...961...67T,Ercolino2025A&A...696A.103E}, potentially contributing to the diversity shown in Figure~\ref{fig:color_fwhm_comp}. In addition, in the binary scenario, the CSM may form a circumbinary disk around the progenitor system \citep[e.g.,][]{Lu23}. Such an asymmetric structure of CSM may lead to different observables from different viewing angles. \corr{For example, in the presence of a circumstellar disk, the viewing angle may significantly affect the peak luminosity of Type Ibn SNe \citep[e.g.,][]{Suzuki2019ApJ...887..249S}, as well as other features such as the strength of O~I lines in late-time spectra \citep{Dong2024ApJ...977..254D}. }

\section{Summary and Conclusions} \label{sec:conclusions}
By comparing with other Type Ibn SNe, we find that SN~2024acyl shows the typical photometric and spectroscopic features of the class. However, the large host offset and low SFR at the explosion site of SN~2024acyl are unusual for a massive star origin. Along with the atypical host environments of PS1-12sk and SN~2023tsz, this raises the question of whether all Type Ibn SNe originate from massive stars. 

We also identify a spectral diversity among Type Ibn SNe around 20--35 days after peak brightness. 
We found that, based on the spectral and photometric properties at this phase range, Type Ibn SNe can be separated into two groups: Group I, which exhibits bluer color ($g-i < 0.2$) and narrower $\rm He~I\,\lambda7065$ lines, and Group II, which shows redder color ($g-i > 0.2$) and broader $\rm He~I\,\lambda7065$ lines. We also find that Group I objects tend to have higher peak luminosities. At very late times ($\sim$80 days after peak), Group I objects display strong, narrow He emission lines, while Group II objects show broader He lines along with features from intermediate-mass elements such as calcium and oxygen.  

This diversity connects closely to the early-time classification noted in H17: specifically, Group I objects show He~I emission lines around peak, while Group II objects exhibit narrow P-Cygni profiles. Our analysis confirms the presence of spectral diversity in Type Ibn SNe and introduces a more quantitative way for describing this diversity. \corr{Based on the current dataset, we find no clear connection between this spectral diversity and either the host environments of Type Ibn SNe or their pre-explosion activity.}

This spectral diversity is likely driven by differences in CSM configuration and/or explosion energy. Group I objects may be associated with denser and more extended CSM and/or lower explosion energies, such that the ejecta are efficiently decelerated and broad lines do not emerge. In contrast, Group II objects may have less dense and less extended CSM and/or higher explosion energies, allowing the ejecta to overtake the CSM quickly and reveal broad lines from the inner ejecta. \corr{These differences may result from a range of progenitor properties, such as different helium star mass, orbital period and companion type if they are in binary systems, and may further imply there are diverse progenitor channels for Type Ibn SNe.}
However, \corr{based on current data we have}, we cannot rule out the possibility that there is a continuum between these two groups and further observational evidence is required.

The observed diversity in both host environments and spectral properties of Type Ibn SNe may point toward multiple progenitor channels for Type Ibn SNe. A systematic study of host environments will be crucial to test this possibility.
\corr{In addition, a systematic analysis of flash spectroscopy is needed to investigate whether the two groups differ in other properties, such as the chemical composition of the surrounding CSM.}
Future efforts combining X-ray, UV, and radio observations with optical spectroscopy from early to late times will be needed to disentangle whether the two groups we identified represent intrinsically distinct progenitor channels or instead a continuum shaped by explosion energy, CSM geometry, and the amount of CSM. Moreover, with the depth of the upcoming Vera C. Rubin Observatory's Legacy Survey of Space and Time (LSST), pre-explosion activity of Type Ibn SNe can be systematically probed out to much greater distances, providing a complete picture on their mass-loss history and the physical processes that lead to these explosions.

\section*{Acknowledgements}
We acknowledge Harsh Kumar for help with the KeplerCam reductions.

The Villar Astro Time Lab acknowledges support through the David and Lucile Packard Foundation, the Research Corporation for Scientific Advancement (through a Cottrell Fellowship), the National Science Foundation under AST-2433718, AST-2407922 and AST-2406110, as well as an Aramont Fellowship for Emerging Science Research.
C.G. is supported by a VILLUM FONDEN Villum Experiment grant (VIL69896).
A.S. is is supported by a VILLUM FONDEN Villum Experiment grant (VIL69896) and a research grant (VIL54489) from VILLUM FONDEN.
S.V.\ and the UC Davis time-domain research team acknowledge support by NSF grants AST-2407565.
R.J.F.\ is supported in part by a fellowship from the David and Lucile Packard Foundation.
D.T.\ is supported by the Sherman Fairchild Postdoctoral Fellowship in Caltech.
This publication was made possible through the support of an LSST-DA Catalyst Fellowship to K.A.B, funded through Grant 62192 from the John Templeton Foundation to LSST Discovery Alliance. The opinions expressed in this publication are those of the authors and do not necessarily reflect the views of LSST-DA or the John Templeton Foundation.
This material is based upon work supported by the National Science Foundation Graduate Research Fellowship Program under Grant Nos.\ 1842402 and 2236415. Any opinions, findings, conclusions, or recommendations expressed in this material are those of the authors and do not necessarily reflect the views of the National Science Foundation.
D.O.J.\ acknowledges support from NSF grants AST-2407632, AST-2429450, and AST-2510993, NASA grant 80NSSC24M0023, and HST/JWST grants HST-GO-17128.028 and JWST-GO-05324.031, awarded by the Space Telescope Science Institute (STScI), which is operated by the Association of Universities for Research in Astronomy, Inc., for NASA, under contract NAS5-26555.
Q.W. is supported by the Sagol Weizmann-MIT Bridge Program.

This work makes use of data from the Las Cumbres Observatory global telescope network. The LCO team is supported by NSF grants AST-2308113 and AST-1911151.

This work has made use of data from the Asteroid Terrestrial-impact Last Alert System (ATLAS) project. The Asteroid Terrestrial-impact Last Alert System (ATLAS) project is primarily funded to search for near earth asteroids through NASA grants NN12AR55G, 80NSSC18K0284, and 80NSSC18K1575; byproducts of the NEO search include images and catalogs from the survey area. This work was partially funded by Kepler/K2 grant J1944/80NSSC19K0112 and HST GO-15889, and STFC grants ST/T000198/1 and ST/S006109/1. The ATLAS science products have been made possible through the contributions of the University of Hawaii Institute for Astronomy, the Queen’s University Belfast, the Space Telescope Science Institute, the South African Astronomical Observatory, and The Millennium Institute of Astrophysics (MAS), Chile.

This work includes data obtained with the Swope telescope at Las Campanas Observatory, Chile, as part of the Swope Time Domain Key Project (PI A.\ Piro; CoIs Burns, Coulter, Cowperthwaite, Dimitriadis, Drout, Foley, French, Holoien, Hsiao, Kilpatrick, Madore, Phillips, Rojas-Bravo).

Based on observations obtained at the international Gemini Observatory, a program of NSF's NOIRLab, which is managed by the Association of Universities for Research in Astronomy (AURA) under a cooperative agreement with the National Science Foundation. On behalf of the Gemini Observatory partnership: the National Science Foundation (United States), National Research Council (Canada), Agencia Nacional de Investigaci\'{o}n y Desarrollo (Chile), Ministerio de Ciencia, Tecnolog\'{i}a e Innovaci\'{o}n (Argentina), Minist\'{e}rio da Ci\^{e}ncia, Tecnologia, Inova\c{c}\~{o}es e Comunica\c{c}\~{o}es (Brazil), and Korea Astronomy and Space Science Institute (Republic of Korea).

Some of the data presented herein were obtained at Keck Observatory, which is a private 501(c)3 non-profit organization operated as a scientific partnership among the California Institute of Technology, the University of California, and the National Aeronautics and Space Administration. The Observatory was made possible by the generous financial support of the W.\ M.\ Keck Foundation. 

This work was enabled by observations made from the Gemini North and Keck telescopes, located within the Maunakea Science Reserve and adjacent to the summit of Maunakea.  The authors wish to recognize and acknowledge the very significant cultural role and reverence that the summit of Maunakea has always had within the indigenous Hawaiian community.  We are most fortunate to have the opportunity to conduct observations from this mountain.

We acknowledge the use of public data from the {\it Swift} data archive.

Pan-STARRS is a project of the Institute for Astronomy of the University of Hawaii, and is supported by the NASA SSO Near Earth Observation Program under grants 80NSSC18K0971, NNX14AM74G, NNX12AR65G, NNX13AQ47G, NNX08AR22G, 80NSSC21K1572, and by the State of Hawaii.  The Pan-STARRS1 Surveys (PS1) and the PS1 public science archive have been made possible through contributions by the Institute for Astronomy, the University of Hawaii, the Pan-STARRS Project Office, the Max-Planck Society and its participating institutes, the Max Planck Institute for Astronomy, Heidelberg and the Max Planck Institute for Extraterrestrial Physics, Garching, The Johns Hopkins University, Durham University, the University of Edinburgh, the Queen's University Belfast, the Harvard-Smithsonian Center for Astrophysics, the Las Cumbres Observatory Global Telescope Network Incorporated, the National Central University of Taiwan, STScI, NASA under grant NNX08AR22G issued through the Planetary Science Division of the NASA Science Mission Directorate, NSF grant AST-1238877, the University of Maryland, Eotvos Lorand University (ELTE), the Los Alamos National Laboratory, and the Gordon and Betty Moore Foundation.

The Infrared Telescope Facility is operated by the University of Hawaii under contract 80HQTR24DA010 with the National Aeronautics and Space Administration.

Observations reported here were obtained at the MMT Observatory, a joint facility of the Smithsonian Institution and the University of Arizona.

The Young Supernova Experiment (YSE) and its research infrastructure is supported by the European Research Council under the European Union's Horizon 2020 research and innovation programme (ERC Grant Agreement 101002652, PI K.\ Mandel), the Heising-Simons Foundation (2018-0913, PI R.\ Foley; 2018-0911, PI R.\ Margutti), NASA (NNG17PX03C, PI R.\ Foley), NSF (AST--1720756, AST--1815935, PI R.\ Foley; AST--1909796, AST-1944985, PI R.\ Margutti), the David \& Lucille Packard Foundation (PI R.\ Foley), VILLUM FONDEN (project 16599, PI J.\ Hjorth), and the Center for AstroPhysical Surveys (CAPS) at the National Center for Supercomputing Applications (NCSA) and the University of Illinois Urbana-Champaign.

YSE-PZ was developed by the UC Santa Cruz Transients Team with support from The UCSC team is supported in part by NASA grants NNG17PX03C, 80NSSC19K1386, and 80NSSC20K0953; NSF grants AST-1518052, AST-1815935, and AST-1911206; the Gordon \& Betty Moore Foundation; the Heising-Simons Foundation; a fellowship from the David and Lucile Packard Foundation to R.\ J.\ Foley; Gordon and Betty Moore Foundation postdoctoral fellowships and a NASA Einstein fellowship, as administered through the NASA Hubble Fellowship program and grant HST-HF2-51462.001, to D.~O.~Jones; and a National Science Foundation Graduate Research Fellowship, administered through grant No.\ DGE-1339067, to D.~A.~Coulter.

This research has made use of the NASA/IPAC Extragalactic Database (NED), which is funded by the National Aeronautics and Space Administration and operated by the California Institute of Technology.

This research made use of Photutils, an Astropy package for detection and photometry of astronomical sources \citep{larry_bradley_2022_6825092}.

\facilities{ADS, ATLAS, LCOGT (SBIG, Sinistro, FLOYDS), Gemini:North (GMOS), Keck:I (LRIS), Keck:II (MOSFIRE), NED, Swift (UVOT), LBT (MODS), Swope, FLWO:1.5m, FLWO:1.2m, TESS, ATLAS, IRTF
}

\software{Astropy \citep{astropy13,astropy18, Astropy2022ApJ...935..167A}, 
emcee \citep{foreman-mackey_emcee_2013}
          HOTPANTS \citep{Becker2015},
          Matplotlib \citep{Hunter2007},
          NumPy \citep{2020Natur.585..357H},
          Prospector \citep{Johnson2021ApJS..254...22J},
          PYRAF \citep{2012ascl.soft07011S},
          Pandas \citep{mckinney-proc-scipy-2010},
          SciPy \citep{2020NatMe..17..261V},
          SWarp \citep{Bertin2002},
          HOTPANTS \citep{Becker2015},
          LCOGTSNpipe \citep{Valenti2016}, 
          Light Curve Fitting \citep{griffin_hosseinzadeh_2020_4312178},
          LPipe \citep{Perley2019PASP..131h4503P},
          YSE-PZ \citep{2022Coulter_YSEPZ,2023PASP_YSEPZ}
          }

\appendix
\twocolumngrid

\section{Light Curve Modeling} \label{append:lc_fit}
\begin{figure}
\centering
\includegraphics[width=1\linewidth]{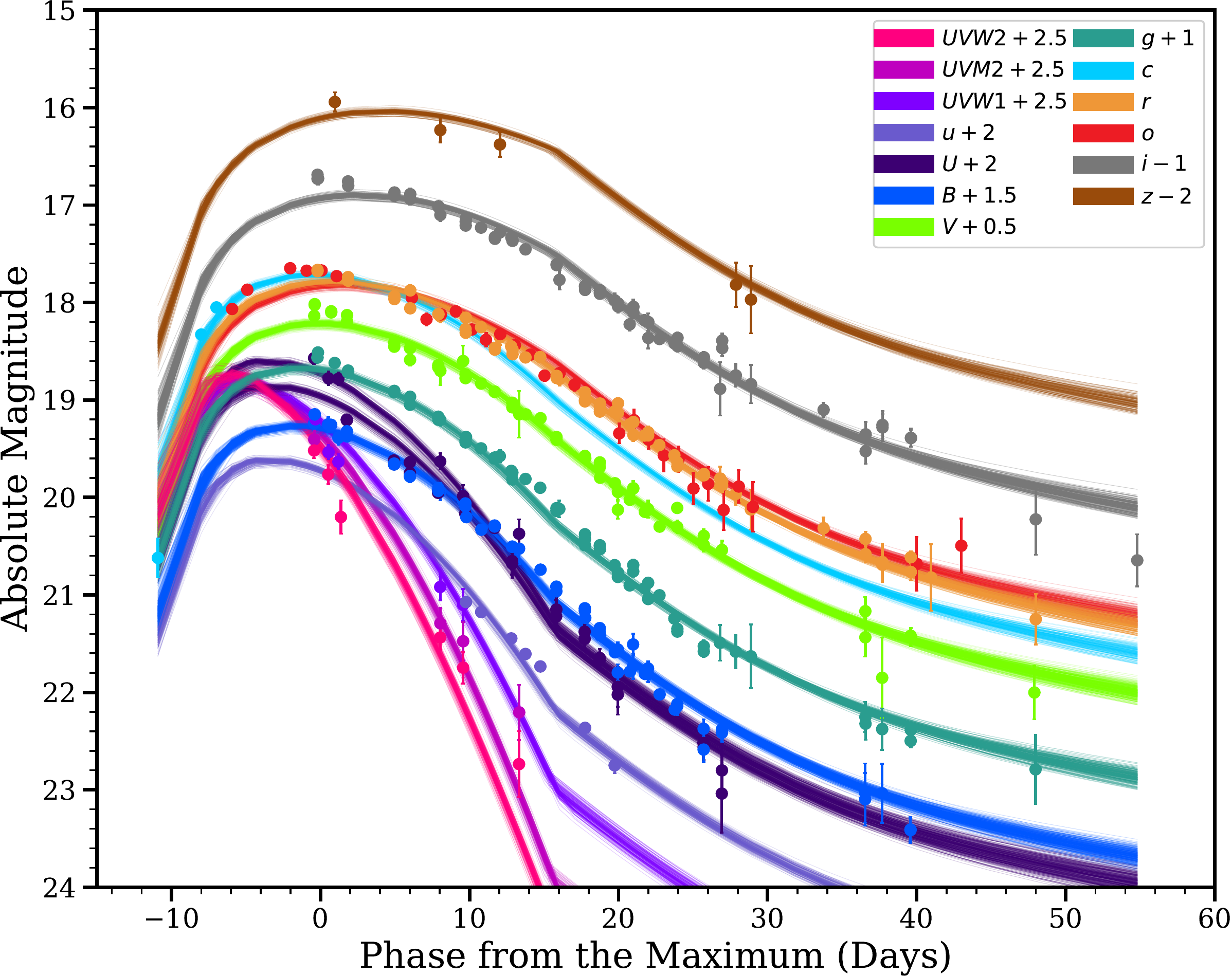}
\caption{Multiband light curve modeling using MOSF{\sc i}T.
\label{fig:mosfit}}
\end{figure}

We fit the multiband light curves with a CSM+Ni model using MOSFiT \citep{Guillochon2018ApJS..236....6G}. We fix the optical opacity ($\kappa$) to be 0.1~$\rm cm^{2}\,g^{-1}$, $n=10$, $s=0$, and $\delta=1$ following \cite{Pellegrino2022ApJ...926..125P}. The $\gamma$-ray opacity is fixed to 0.03 $\rm cm^{2}\,g^{-1}$ following \cite{Valenti2008}. The conversion efficiency of the shock kinetic energy to radiation ($\epsilon$) is set to 0.5. The free parameters in our fit include the nickel fraction of the ejecta mass ($f_{\rm Ni}$), the CSM mass ($M_{\rm CSM}$), the ejecta mass ($M_{ej}$), the inner radius of CSM ($R_{0}$), the CSM density at $R_{0}$ ($\rho_{\rm CSM}$), the minimum temperature of the photosphere before it starts to recede ($T_{\rm min}$), explosion epoch relative to the first detection ($t_{\rm exp}$), and the ejecta velocity ($v_{ej}$). For a detailed description of the model and its parameters, we refer the reader to \cite{Arnett1982}, \cite{Chatzopoulos2013ApJ...773...76C}, and \cite{Villar2017ApJ...849...70V}. The priors and best-fit values are listed in Table~\ref{tab:mosfit}, and the light curve fit is shown in Figure~\ref{fig:mosfit}. We note that the models used here are likely oversimplified, and the derived parameters should be regarded as order-of-magnitude estimates.


\begin{deluxetable}{cccc}
\tablenum{A1}
\tablecaption{The priors and the best-fit parameters of the CSM+Ni model \label{tab:mosfit}}
\tablewidth{0pt}
\tablehead{
\colhead{Parameter} & \colhead{Unit} & \colhead{Prior} & \colhead{Best Value}  
}
\startdata 
$f_{\rm Ni}$ & - & ($10^{-5}$, 1.0) & $0.04^{+0.01}_{-0.01}$ \\
$M_{\rm CSM}$ & $\rm M_{\odot}$ & ($10^{-3}$, 30.0) & $0.36^{+0.04}_{-0.03}$ \\
$M_{\rm ej}$ & $\rm M_{\odot}$ & (0.1, 30.0) & $0.70^{+0.20}_{-0.10}$ \\
$R_{0}$ & $\rm 10^{14}\,cm$ & (0.1, 300) & $1.62^{+2.30}_{-2.76}$ \\
$\rho_{\rm CSM}$ & $10^{-15}\rm \,g\,cm^{-3}$ & ($10^{-17}$, $10^{-10}$) & $8128.31^{+1684.46}_{-1310.14}$  \\
$T_{\rm min}$ & $\rm K$ & ($10^3$, $5\times10^{4}$) & $6456.54^{+148.67}_{-148.67}$  \\
$t_{\rm exp}$ & days & (-3, 0) & $-2.90^{+0.15}_{-0.08}$  \\
$v_{ej}$ & $\rm km\,s^{-1}$ & ($5\times10^{3}$, $10^{4}$) & $5113.04^{+85.47}_{-48.67}$ \\
\enddata{}
\end{deluxetable}

\section{Data Reduction} \label{append:data_reduc}
\subsection{Photometry Reduction}
Images obtained through the Las Cumbres Observatory were preprocessed by BANZAI \citep{McCully2018SPIE10707E..0KM}. The PSF photometry was then extracted using the PyRAF-based photometric reduction pipeline {\sc lcogtsnpipe} \citep{Valenti2016}. 
We calibrated the $gri$ magnitudes to the APASS catalog \citep{Henden2018AAS...23222306H}, and we calibrated $UBV$ magnitudes to the Landolt catalog \citep{Landolt1992AJ....104..340L}.

$griz$-band images obtained via the KeplerCam instrument on
the FLWO 1.2-m telescope at the Fred Lawrence Whipple Observatory between 2024 December 9 and 2025 Feburary 27 was reduced using a custom pipeline. Images taken from the same night were first stacked using Swarp \cite{Bertin2002}. The PSF photometry were extracted using SExtractor \citep{Bertin1996A&AS..117..393B} and calibrated to the Pan-STARRS catalog \citep{Chambers2016arXiv161205560C}. Image subtraction was done using Pyzogy \citep{Guevel2017zndo...1043973G}.

SN~2024acyl was observed by UVOT instrument on the Neil Gehrels $Swift$ Observatory \citep{Gehrels2004} between 2024 December 18 and 2025 January 1. For the Swift images, we performed aperture photometry with an aperture size of 3\arcsec at the position of SN~2024acyl on \textit{Swift} UVOT images using the High-Energy Astrophysics software (HEA-Soft). Background variations in individual images were removed using a 5\arcsec aperture placed on a blank section of the sky. Zero-points were chosen from \cite{Breeveld2011} with time-dependent sensitivity corrections updated in 2020.

The Gemini-North GMOS $gri$-band images were taken on 2025 February 22, through program GN-2025A-A-208. The images were reduced using the Gemini IRAF package \citep{GeminiIRAF2016ascl.soft08006G}, a software suite built on IRAF \citep{Tody1986SPIE..627..733T, Tody1993ASPC...52..173T} and designed to process data taken with the Gemini telescopes. PSF photometry was extracted using Photutils \citep{larry_bradley_2022_6825092} and calibrated to the Pan-STARRS catalog \citep{Chambers2016arXiv161205560C}.

We observed SN~2024acyl through the 1-meter Henrietta Swope telescope from 2024 December 19 to 2025 January 12 in the $uBVgri$ bands. The images were reduced using the Photpipe pipeline \citep{Rest2005ApJ...634.1103R, Rest2014ApJ...795...44R}. The complete data reduction procedure is described in \citet{Kilpatrick_2018}.

The TESS data between JD~2460610.56--2460636.05 were reduced with the \textsc{TESSreduce} \cite{RiddenHarper2021arXiv211115006R} package.

\subsection{Spectra Reduction}
The FLOYDS spectra were reduced following standard procedures using the FLOYDS pipeline \citep{Valenti2014}.

The MMT Binospec spectra were obtained with the 270 line grating, LP3800 blocking filter, a 1" slit width, and a central wavelength of 6500~$\rm \AA$. The spectra were reduced in the standard manner with {\sc PypeIt} \citep{pypeit:joss_arXiv,pypeit:joss_pub,pypeit:zenodo}.

All LRIS spectra were obtained using the D560 dichroic, the 600/4000 grism, and the 400/8500 grating with a slit width of 1" and a central wavelength of 7770~$\rm \AA$.
The spectra were reduced in a standard way using the LPipe pipeline \citep{Perley2019PASP..131h4503P}.

The FAST spectrum is reduced using the \texttt{IRAF} package \citep{Tody1986SPIE..627..733T,Tody1993ASPC...52..173T}.

For the Gemini-N GMOS spectrum taken on 2025 February 22 (76 days after maximum), no trace is visible in the 2D image, as the transient was already too faint. A blind offset was performed during the observation, placing a nearby offset star (Gaia DR3 128634717191988736) on the slit. To extract the trace at the position of SN~2024acyl, we first extracted the trace of the offset star and then applied the same trace at the SN position. Both spectra were flux calibrated using the standard star. To verify the flux calibration, synthetic photometry of the offset star spectrum was compared to its Gaia DR3 catalog magnitude.

The IRTF SpeX NIR data obtained in an ABBA dithering pattern were reduced using a standard implementation of Spextool
\citep{Cushing2004PASP116362C}. The output was telluric corrected using a standard A0V star observed at similar airmass adjacent to the science target, following the prescription described in \citep{Vacca2003PASP115389V}. This spectrum has a low S/N, so it is not used for analysis.

The Keck~II MOSFIRE spectrum was reduced in the standard manner with {\sc PypeIt} \citep{pypeit:joss_arXiv,pypeit:joss_pub,pypeit:zenodo}. This spectrum has a low S/N, so it is not used for analysis.

\section{Spectroscopic Observations}\label{appendix:spec_table}
Table \ref{tab:spectra} shows a log of the spectroscopic observations of SN~2024acyl and SN~2023xgo. 

\begin{deluxetable*}{cccccc}
\tablenum{A1}
\tablecaption{Spectroscopic observations of SN~2024acyl and SN~2023xgo\label{tab:spectra}}
\tablewidth{0pt}
\tablehead{
\colhead{Object} & \colhead{UT Date} & \colhead{Julian Date (Days)} & \colhead{Phase (Days)} & \colhead{Telescope} & \colhead{Instrument} 
}
\startdata 
SN~2024acyl & 2024-12-04 & 2460649.3 & -4.5 & - & ALPY200\\
SN~2024acyl & 2024-12-06 & 2460650.6 & -3.2 & NTT & EFOSC2\\
SN~2024acyl & 2024-12-07 & 2460651.8 & -2.0 & MMT & BinoSpec\\
SN~2024acyl & 2024-12-08 & 2460652.9 & -0.9 & FTN & FLOYDS\\
SN~2024acyl & 2024-12-09 & 2460653.6 & -0.1 & NTT & EFOSC2\\
SN~2024acyl & 2024-12-10 & 2460654.9 & 1.1 & FTN & FLOYDS\\
SN~2024acyl & 2024-12-17 & 2460661.7 & 7.9 & FTN & FLOYDS\\
SN~2024acyl & 2024-12-18 & 2460662.9 & 9.1 & FTN & FLOYDS\\
SN~2024acyl & 2024-12-19 & 2460663.9 & 10.1 & FTN & FLOYDS\\
SN~2024acyl & 2024-12-20 & 2460664.9 & 11.1 & FTN & FLOYDS\\
SN~2024acyl & 2024-12-21 & 2460665.9 & 12.1 & FTN & FLOYDS\\
SN~2024acyl & 2024-12-22 & 2460666.8 & 13.0 & FTN & FLOYDS\\
SN~2024acyl & 2024-12-24 & 2460668.9 & 15.1 & FTN & FLOYDS\\
SN~2024acyl & 2024-12-26 & 2460670.7 & 16.9 & MMT & BinoSpec\\
SN~2024acyl & 2024-12-27 & 2460671.9 & 18.1 & IRTF & Spex$\ddag$ \\
SN~2024acyl & 2025-01-02 & 2460677.9 & 24.1 & Keck-I & LRIS\\
SN~2024acyl & 2025-01-07 & 2460682.7 & 28.9 & MMT & BinoSpec\\
SN~2024acyl & 2025-01-28 & 2460703.7 & 49.9 & MMT & BinoSpec\\
SN~2024acyl & 2025-02-16 & 2460722.8 & 69.0 & Keck-II & MOSFIRE$\ddag$\\
SN~2023xgo & 2023-12-12 & 2460290.5 & 27.6 & Keck-I & LRIS\\
\enddata{}
\tablecomments{A log of the spectroscopic observations.}
\tablenotetext{$\ddag$}{These spectra are too noisy to be used in the analysis and are shown in Figure~\ref{fig:nir} for completeness only.}
\end{deluxetable*}

\begin{figure}
\centering
\includegraphics[width=1\linewidth]{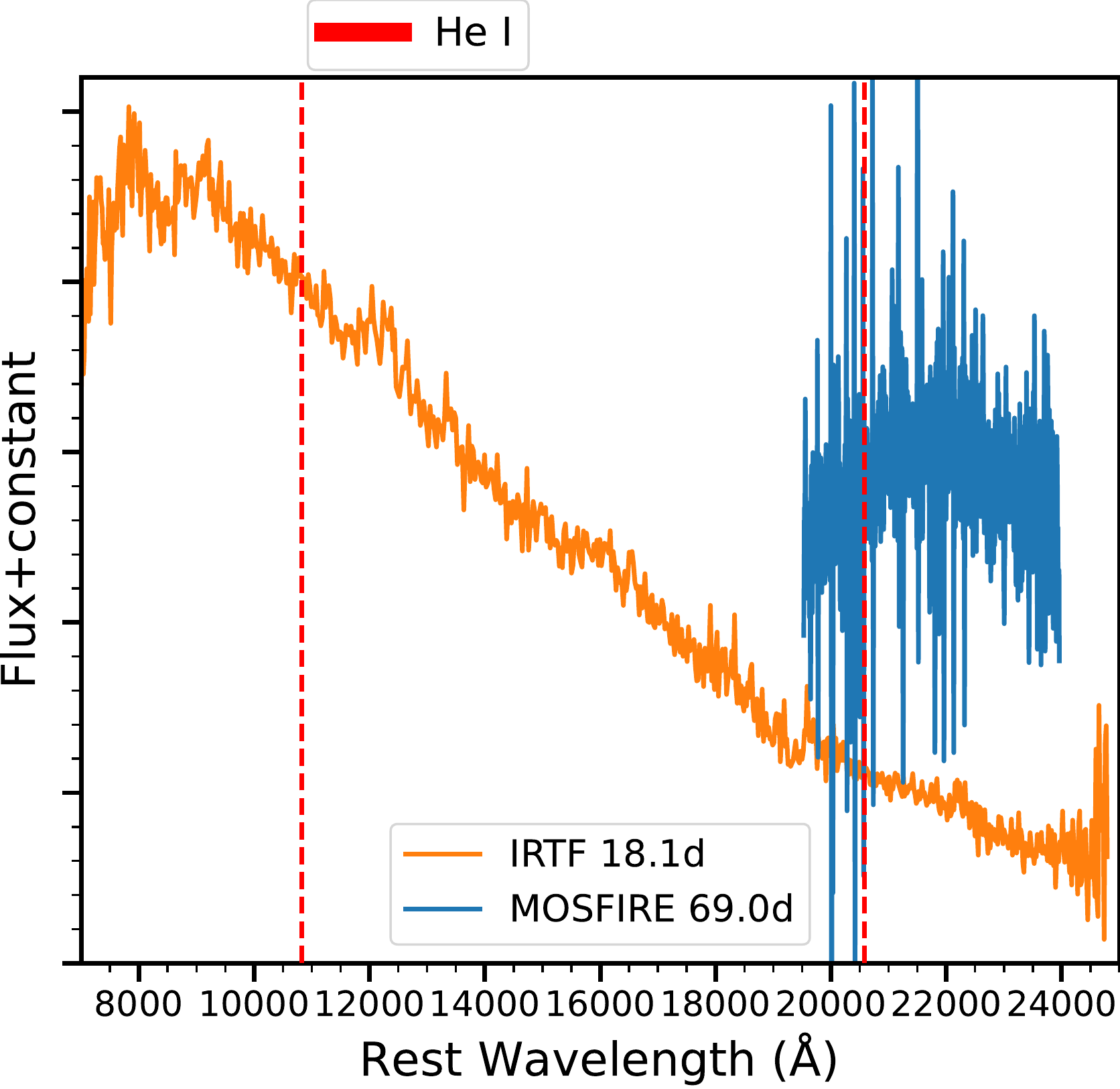}
\caption{The NIR spectra of SN~2024acyl taken with IRTF Spex and Keck-II MOSFIRE. The red dashed lines indicate the position of He~I lines.
\label{fig:nir}}
\end{figure}

\bibliography{SN2024acyl}{}
\bibliographystyle{aasjournal}



\end{CJK*}
\end{document}